\documentclass[]{aastex631}

\usepackage{float}
\usepackage[caption=false]{subfig}

\usepackage{graphicx}

\begin{document}
\title{Simons Observatory: Characterization of the Large Aperture Telescope Receiver }

\author[0000-0002-2971-1776]{Tanay Bhandarkar}
\affiliation{Department of Physics and Astronomy, University of Pennsylvania, Philadelphia, PA, 19104, USA}
\author[0000-0001-6519-502X]{Saianeesh K. Haridas}
\affiliation{Department of Physics and Astronomy, University of Pennsylvania, Philadelphia, PA, 19104, USA}

\author{Jeff Iuliano}
\affiliation{Department of Physics and Astronomy, University of Pennsylvania, Philadelphia, PA, 19104, USA}
\author[0000-0001-5374-1767]{Anna Kofman}
\affiliation{Department of Physics and Astronomy, University of Pennsylvania, Philadelphia, PA, 19104, USA}
\affiliation{Department of Astronomy and Astrophysics, University of Chicago, Chicago, IL, 60637, USA}
\author[0000-0003-4629-5759]{Alex Manduca}
\affiliation{Department of Physics and Astronomy, University of Pennsylvania, Philadelphia, PA, 19104, USA}
\author{Karen Perez Sarmiento}
\affiliation{Department of Physics and Astronomy, University of Pennsylvania, Philadelphia, PA, 19104, USA}
\author[0000-0003-1842-8104]{John Orlowski-Scherer}
\affiliation{Department of Physics and Astronomy, University of Pennsylvania, Philadelphia, PA, 19104, USA}
\author[0000-0002-6452-4220]{Thomas P. Satterthwaite}
\affiliation{Department of Physics, Stanford University, Stanford, CA 94305, USA}
\affiliation{Kavli Institute for Particle Astrophysics and Cosmology, Stanford, CA 94305 USA}
\author[0000-0002-8710-0914]{Yuhan Wang}
\affiliation{Department of Physics, Cornell University, Ithaca, NY 14853, USA}
\author[0000-0002-9957-448X]{Zeeshan Ahmed}
\affiliation{Kavli Institute for Particle Astrophysics and Cosmology, Stanford, CA 94305 USA}
\affiliation{SLAC National Accelerator Laboratory , Menlo Park, California 94025, USA}
\author[0000-0002-6338-0069]{Jason E. Austermann}
\affiliation{National Institute of Standards and Technology, Quantum Sensors Division, 325 Broadway, Boulder, CO 80305, USA}
\author[0000-0002-3376-8660]{Kyuyoung Bae}
\affiliation{Department of Physics, University of Colorado, Boulder, CO, 80309, USA}
\affiliation{National Institute of Standards and Technology, Quantum Sensors Division, 325 Broadway, Boulder, CO 80305, USA}

\author[0000-0002-6362-6524]{Gabriele Coppi}
\affiliation{Department of Physics, University of Milano-Bicocca, Milano, Italy}
\affiliation{Istituto Nazionale di Fisica Nucleare, INFN, Sezione Milano-Bicocca, Italy}
\author[0000-0002-3169-9761]{Mark J. Devlin}
\affiliation{Department of Physics and Astronomy, University of Pennsylvania, Philadelphia, PA, 19104, USA}
\author[0000-0002-1940-4289]{Simon R Dicker}
\affiliation{Department of Physics and Astronomy, University of Pennsylvania, Philadelphia, PA, 19104, USA}
\author[0009-0006-8427-6259]{Peter N. Dow}
\affiliation{Department of Astronomy, University of Virginia, 530 McCormick Rd, Charlottesville, VA 22904}
\author[0000-0002-9693-4478]{Shannon M. Duff}
\affiliation{National Institute of Standards and Technology, Quantum Sensors Division, 325 Broadway, Boulder, CO 80305, USA}
\author[0000-0002-9962-2058]{Daniel Dutcher}
\affiliation{Department of Physics, Princeton University, Princeton, NJ, 08540, USA}
\author[0000-0001-7225-6679]{Nicholas Galitzki}
\affiliation{Department of Physics, University of Texas at Austin, Austin, TX, 78712, USA}
\affiliation{Weinberg Institute for Theoretical Physics, Texas Center for Cosmology and Astroparticle Physics, Austin, TX 78712, USA}
\author[0000-0003-1760-0355]{Jon E. Gudmundsson}
\affiliation{Science Institute, University of Iceland, 107 Reykjavik, Iceland; The Oskar Klein Centre, Department of Physics, Stockholm University, AlbaNova, SE-10691 Stockholm, Sweden}
\author[0000-0001-7878-4229]{Shawn W. Henderson}
\affiliation{Kavli Institute for Particle Astrophysics and Cosmology, Stanford, CA 94305 USA}
\affiliation{SLAC National Accelerator Laboratory , Menlo Park, California 94025, USA}
\author[0000-0002-2781-9302]{Johannes Hubmayr}
\affiliation{National Institute of Standards and Technology, Quantum Sensors Division, 325 Broadway, Boulder, CO 80305, USA}
\author[0000-0002-6898-8938]{Bradley R. Johnson}
\affiliation{Department of Astronomy, University of Virginia, 530 McCormick Rd, Charlottesville, VA 22904}
\author[0000-0003-1401-8415]{Matthew A. Koc}
\affiliation{National Institute of Standards and Technology, Quantum Sensors Division, 325 Broadway, Boulder, CO 80305, USA}
\affiliation{Department of Physics, University of Colorado, Boulder, CO, 80309, USA}
\author[0000-0003-0744-2808]{Brian J. Koopman}
\affiliation{Wright Laboratory, Department of Physics, Yale University, New Haven, Connecticut 06511, USA}
\author[0000-0002-5900-2698]{Michele Limon}
\affiliation{Department of Physics and Astronomy, University of Pennsylvania, Philadelphia, PA, 19104, USA}
\author[0000-0003-2381-1378]{Michael J Link}
\affiliation{National Institute of Standards and Technology, Quantum Sensors Division, 325 Broadway, Boulder, CO 80305, USA}
\author[0000-0001-7694-1999]{Tammy J. Lucas}
\affiliation{National Institute of Standards and Technology, Quantum Sensors Division, 325 Broadway, Boulder, CO 80305, USA}
\author[0000-0002-7340-9291]{Jenna E. Moore}
\affiliation{Department of Physics, Duke University, Durham, NC, 27708, USA}
\author[0000-0002-8307-5088]{Federico Nati}
\affiliation{Department of Physics, University of Milano-Bicocca, Milano, MI, 20126, Italy}
\author[0000-0001-7125-3580]{Michael D. Niemack}
\affiliation{Department of Physics, Cornell University, Ithaca, NY 14853, USA}
\author[0000-0002-9246-5571]{Carlos E. Sierra}
\affiliation{Kavli Institute for Particle Astrophysics and Cosmology, Stanford, CA 94305 USA}
\affiliation{Department of Physics, Stanford University, Stanford, CA 94305, USA}
\author[0000-0001-7480-4341]{Max Silva-Feaver}
\affiliation{Wright Laboratory, Department of Physics, Yale University, New Haven, Connecticut 06511, USA}
\author[0000-0003-2893-9039]{Robinjeet Singh}
\affiliation{Department of Physics, University of Colorado, Boulder, CO, 80309, USA}
\affiliation{National Institute of Standards and Technology, Quantum Sensors Division, 325 Broadway, Boulder, CO 80305, USA}
\author[0000-0002-7020-7301]{Suzanne T. Staggs}
\affiliation{Department of Physics, Princeton University, Princeton, NJ, 08540, USA}
\author[0000-0002-1187-9781]{Rita F. Sonka}
\affiliation{Department of Physics, Princeton University, Princeton, NJ, 08540, USA}
\author{Robert J. Thornton}
\affiliation{Department of Physics, West Chester University of Pennsylvania, West Chester, PA 19383, USA}
\affiliation{Department of Physics and Astronomy, University of Pennsylvania, 209 South 33rd Street, Philadelphia, PA 19104, USA}
\author[0000-0002-1667-2544]{Tran Tsan}
\affiliation{Physics Division, Lawrence Berkeley National Laboratory, Berkeley, CA 94720, USA}
\author{Jeff L. Van Lanen}
\affiliation{National Institute of Standards and Technology, Quantum Sensors Division, 325 Broadway, Boulder, CO 80305, USA}
\author[0000-0002-2105-7589]{Eve M. Vavagiakis}
\affiliation{Department of Physics, Duke University, Durham, NC, 27708, USA}
\affiliation{Department of Physics, Cornell University, Ithaca, NY, 14853, USA}
\author[0000-0003-2467-7801]{Michael R Vissers}
\affiliation{National Institute of Standards and Technology, Quantum Sensors Division, 325 Broadway, Boulder, CO 80305, USA}
\author[0000-0002-8309-8298]{Liam Walters}
\affiliation{Department of Astronomy, University of Virginia, 530 McCormick Rd, Charlottesville, VA 22904}
\author[0000-0002-4495-571X]{Mario Zannoni}
\affiliation{Department of Physics, University of Milano-Bicocca, Milano, Italy}
\author[0000-0003-4645-7084]{Kaiwen Zheng}
\affiliation{Department of Physics, Princeton University, Princeton, NJ, 08540, USA}

%1667 K Street NW, Suite 800 \\
%Washington, DC 20006, USA}

%% Note that the \and command from previous versions of AASTeX is now
%% depreciated in this version as it is no longer necessary. AASTeX 
%% automatically takes care of all commas and "and"s between authors names.

%% AASTeX 6.31 has the new \collaboration and \nocollaboration commands to
%% provide the collaboration status of a group of authors. These commands 
%% can be used either before or after the list of corresponding authors. The
%% argument for \collaboration is the collaboration identifier. Authors are
%% encouraged to surround collaboration identifiers with ()s. The 
%% \nocollaboration command takes no argument and exists to indicate that
%% the nearby authors are not part of surrounding collaborations.

%% Mark off the abstract in the ``abstract'' environment. 
\begin{abstract}
The Simons Observatory (SO) is a ground-based cosmic microwave background (CMB) survey experiment that currently consists of three 0.42\,m small-aperture telescopes (SATs) and one 6\,m large-aperture telescope (LAT), located at an elevation of 5200\,m in the Atacama Desert in Chile. At the LAT's focal plane, SO will install $>$62,000 transition-edge sensor detectors across 13 optics tubes (OTs) within the Large Aperture Telescope Receiver (LATR), the largest cryogenic camera ever built to observe the CMB. Here we report on the validation of the LATR in the laboratory and the subsequent dark testing and validation within the LAT. We show that the LATR meets cryogenic, optical, and detector specifications required for high-sensitivity measurements of the CMB. At the time of writing, the LATR is installed in the LAT with six OTs (corresponding to $>$31,000 detectors), and the LAT mirrors and remaining seven OTs are undergoing development.

\end{abstract}

%% Keywords should appear after the \end{abstract} command. 
%% The AAS Journals now uses Unified Astronomy Thesaurus concepts:
%% https://astrothesaurus.org
%% You will be asked to selected these concepts during the submission process
%% but this old "keyword" functionality is maintained in case authors want
%% to include these concepts in their preprints.

%% From the front matter, we move on to the body of the paper.
%% Sections are demarcated by \section and \subsection, respectively.
%% Observe the use of the LaTeX \label
%% command after the \subsection to give a symbolic KEY to the
%% subsection for cross-referencing in a \ref command.
%% You can use LaTeX's \ref and \label commands to keep track of
%% cross-references to sections, equations, tables, and figures.
%% That way, if you change the order of any elements, LaTeX will
%% automatically renumber them.
%%
%% We recommend that authors also use the natbib \citep
%% and \citet commands to identify citations.  The citations are
%% tied to the reference list via symbolic KEYs. The KEY corresponds
%% to the KEY in the \bibitem in the reference list below. 

\section{Introduction} \label{sec:intro}
High precision, sensitive observations of the cosmic microwave background (CMB) radiation have guided the development of cosmological theories, such as $\Lambda$CDM. Previous satellite experiments, such as COBE, WMAP and \textit{Planck}, have conducted all-sky surveys and established the fundamental cosmological parameters that inform these models. Ground-based surveys have built on these measurements, utilizing their improved sensitivities and higher resolutions. Small angular scale ground-based experiments ($\sim$1'), such as the Atacama Cosmology Telescope (ACT) (\cite{Thornton_2016}) and the South Pole Telescope (SPT) (\cite{Carlstrom_2011}), have made sky temperature and polarization maps  with greater angular resolution and sensitivity than the full-sky satellite surveys. These maps have enabled a broad range of scientific results including improved measurements of the lensing power spectrum~(\cite{madhavacheril_2023}), the Sunyaev-Zeldovich (SZ) effect~(\cite{Hilton_2021}, \cite{Bleem_2020}), upper bounds on the sum of the neutrino mass~\citep{planck2018_vi}, and the effective number of relativistic species $N_\mathrm{eff}$ (\cite{Abazajian_2015}). Additionally, large angular scale ground-based experiments ($\sim0.5^\circ$) such as the BICEP and Keck Arrays (\cite{Ade_2022}), the Cosmology Large Angular Scale Surveyor (CLASS) (\cite{Dahal_2022}), and POLARBEAR (\cite{POLARBEAR_2020}) focus on constraining the tensor-to-scalar ratio, $r$, through B-mode polarization measurements. Definitive measurements of $r$ would provide clear evidence of early universe inflation (\cite{Kamionkowski_1997}).\par

\begin{figure}
    \centering
    \includegraphics[width=.7\textwidth]{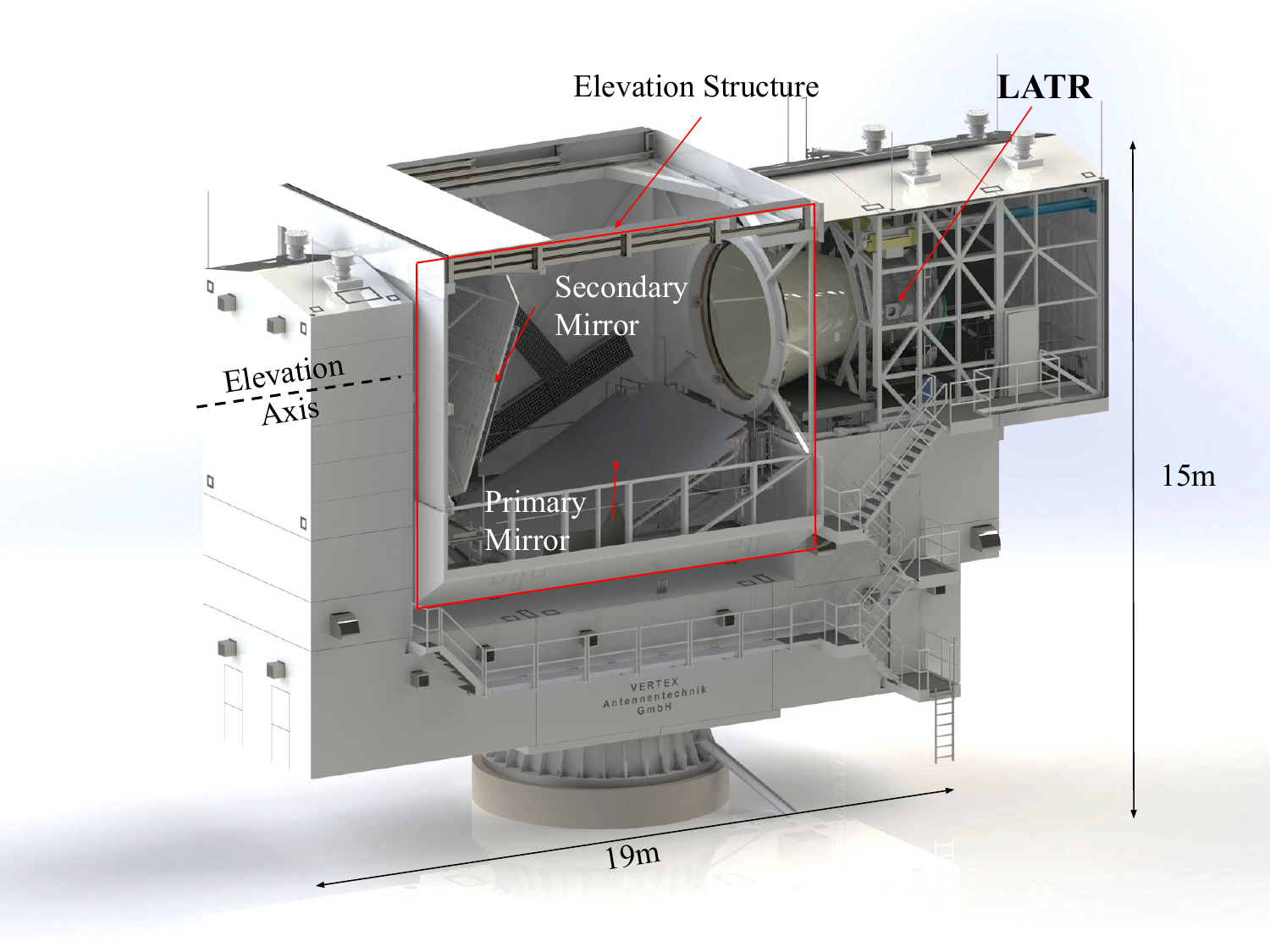}
    \caption{A rendered cross-sectional view of the LAT~\citep{Zhu_2021}. The telescope yoke sits on an azimuth cone (bottom) and supports the elevation structure.  The LATR is located in the receiver cabin (upper right), where it can co-rotate with the elevation structure and collect light reflected from the 5.4\,m secondary mirror (left). See Figure \ref{fig:lat_photo} for a recent photograph of the LAT.}
    \label{fig:lat_cut}
\end{figure}

The Simons Observatory (SO) is a next generation ground-based millimeter wavelength survey located at an elevation of 5200\,meters on Cerro Toco in the Atacama Desert of northern Chile.  It is designed to produce deep temperature and polarization maps of the CMB. The observatory will ultimately consist of six 0.42\,m small-aperture telescopes (SATs, \citep{Galitzki_2024}) and one large-aperture telescope (LAT). The LAT (Figure \ref{fig:lat_cut}) features a crossed-Dragone optical design, with a 6\,m primary mirror and a 5.4\,m secondary mirror. This optical design allows for a large focal plane occupied by the Large Aperture Telescope Receiver (LATR), covering a 6.7$^\circ$ field of view ~\citep{Niemack_2016, Gudmundsson_2021}. At the time of writing, construction of the first three SATs and the LAT is complete.\par % The LATR is designed to house $>$62,000 detectors across 13 modular optics tubes (OT). Optics tube deployment is spread across two phases: the first phase provides resources for seven OTs, and the second phase enables the final six. 
The LAT will survey 40\% of the sky with arcminute resolution, sharing significant overlap with surveys in other wavelengths, such as the Dark Energy Spectroscopic Instrument (DESI) (\cite{DESI_2016}) and the Legacy Survey of Space and Time (LSST) survey of the Vera C. Rubin Observatory (\cite{Ivezić_2019}). The complementary datasets these instruments will produce can be used to constrain the lensing potential and the upper bound of the neutrino mass. Similar to ACT and SPT, the LAT will enable temperature and polarization maps. SO is expected to detect over 16,000 galaxy clusters through the thermal SZ effect, over 24,000 extragalactic point sources \citep{so_forecast_2019}, and a variety of different transient events~\citep{Li2023, Orlowski-Scherer_2024}. \par
In this paper, we start with an overview of the instrument design (Section \ref{sec:LATR_Design}). We then discuss the in-lab cryogenic validation efforts (Section \ref{sec:cryo_validation}) and readout validation tests (Section \ref{sec:rd_validation}). In Section \ref{sec:in_lab}, we describe the instrument's response to different environmental tests. Finally, in Section \ref{sec:LAT} we show initial results of testing conducted during the integration of the telescope and receiver. 

\section{LATR Design} \label{sec:LATR_Design}
 We will briefly review the LATR design here; for a more comprehensive treatment, please refer to \cite{Zhu_2021}. Many of the design details described in the following sections are informed by the experiences of earlier CMB receivers (e.g. ACTPol~\citep{Thornton_2016}). \par
\subsection{Cryogenic and Mechanical Design} \label{sec:Cryomechanical_Design}
\begin{figure}
    \centering
    \includegraphics[width=.7\textwidth]{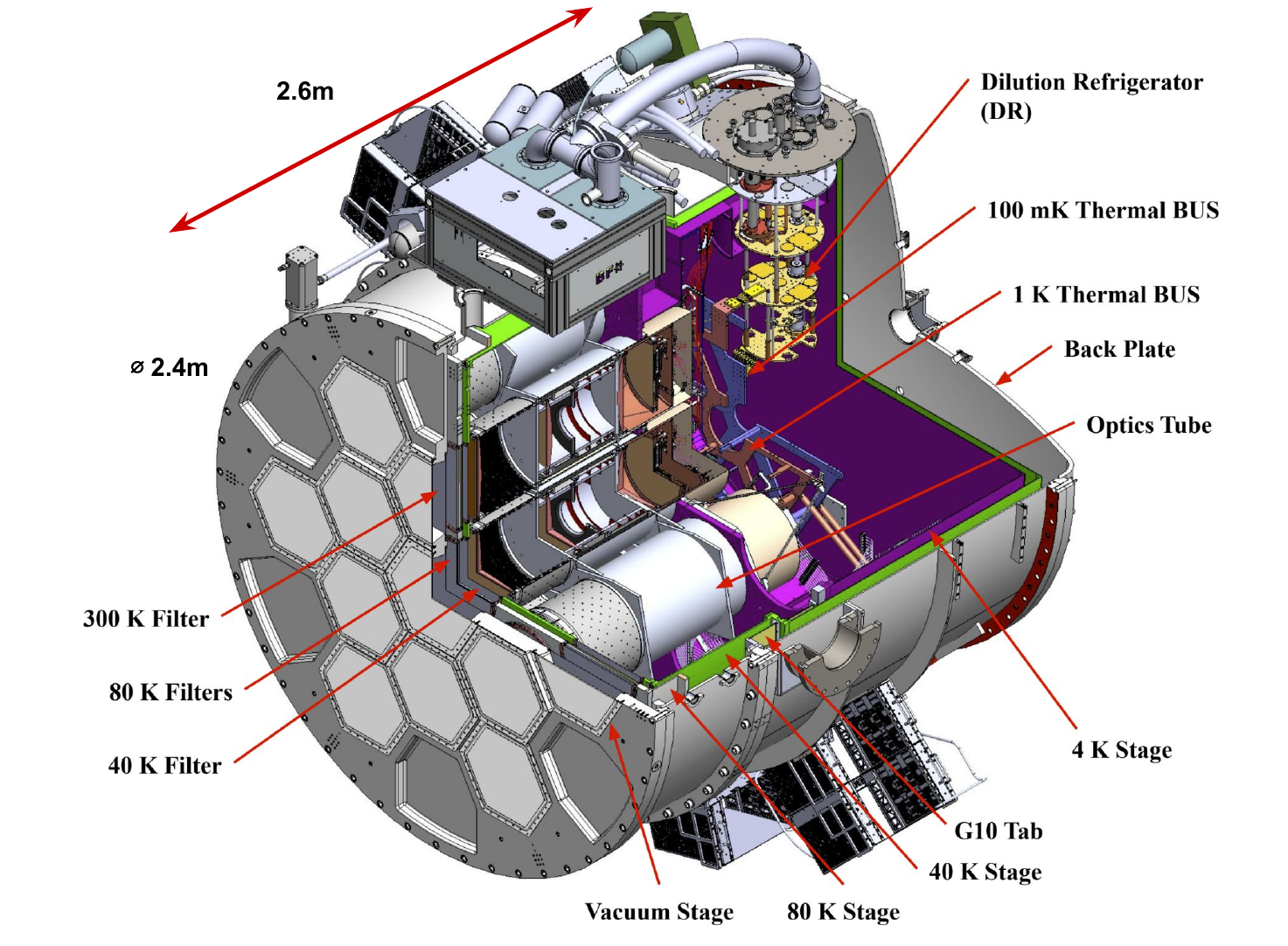}
    \caption{A rendered cross-sectional view of the LATR~\citep{Zhu_2021}. In this figure, we can see the 80\,K stage (grey), 40\,K stage (green), 4\,K stage (purple), 1\,K stage (orange), and 100\,mK stage (indigo). Additionally, we show the cross-section of two OTs, where individual optical components are visible. Light from the secondary LAT mirror enters the cryostat through the hexagonal windows (left) and travels through the OTs to their 100\,mK stage. Finally, warm components, such as the readout electronics and the turbo-pump rack for the dilution refrigerator, are visible on the exterior of the LATR.}
    \label{fig:latr_cut}
\end{figure}
 The LATR consists of five cryogenic stages (80\,K, 40\,K, 4\,K, 1\,K and 100\,mK) and a 300\,K vacuum shell (Figure \ref{fig:latr_cut}). 
 The receiver utilizes superconducting transition edge sensors (TES) designed to operate at 100\,mK (\cite{McCarrick_2021}), which acts as the primary driver of the LATR's cryogenic design. Large temperature variations at the 100\,mK stage can induce thermal noise or even risk heating the TESs out of their transition phase, so a  stable  cold stage is also important for detector performance. \par
 At warmer stages, the desire to balance the LATR's mechanical robustness against its cryogenic requirements inform many of the design decisions for the cryostat. Mechanical and thermal design study simulations were conducted to optimize the LATR's mechanical design~\citep{orlowski_2018}, as were simulations estimating the time to reach the cryogenic operating temperatures~\citep{coppi_2018}.  Minimizing the cooldown time significantly reduces the logistical challenges when operating in the field and increases the on-sky integration time. \par
%The 300\,K front plate holds the first LATR optical elements that light from the LAT's mirrors encounters. 
The 300\,K stage consists of a front plate, a cylindrical vacuum shell, and a dome-shaped back plate. The front plate contains 13 hexagonal cutouts for anti-reflection coated windows, behind which are mounted infrared blocking filters. The vacuum shell contains flanges for vacuum lines and pressure gauges, as well as insertion ports for the cryogenic refrigerators and  readout harnesses. The external ribs of the vacuum shell are also used to mount several components of the readout hardware (Section \ref{sec:warm_readout}). 
The 80\,K stage consists of a 2.1\,m diameter filter plate, and a short 80\,K shell. It is cooled by two PT-90\footnote{https://bluefors.com/products/cryomech-products/} pulse tube cryocoolers, which are coupled to the filter plate via oxygen-free high conductivity (OFHC) copper straps, manufactured by Technology Applications, Inc.\footnote{https://www.techapps.com/} In order to reduce the total thermal load at 80\,K, the entire stage is wrapped in 30 layers of multi-layer insulation (MLI). The 80\,K stage mounts to the 300\,K vacuum shell via a ring of G10 tabs.  \par
The 40\,K stage (highlighted in green in Figure~\ref{fig:latr_cut}) consists of a filter plate, an extended shell, and a back lid. As with the 80\,K stage, it is also mounted to the inside of the 300\,K vacuum shell through a ring of G10 tabs and is wrapped with 30 layers of MLI. The 4\,K stage (highlighted in purple in Figure~\ref{fig:latr_cut}) consists of a filter plate, a radiation shell, and a back lid. It is mounted directly to the 40\,K stage through a second ring of G10 tabs. While the optics tubes (OTs) mount directly to the 4\,K plate, the majority of their 4\,K mass protrudes into the 40\,K cavity. To minimize radiative heating, we wrap these 4\,K sections of the OT in aluminized Mylar. The 40\,K and 4\,K stages are cooled by two dual-stage PT-420$^1$ pulse tube cryocoolers, with each stage coupled to the respective pulse tube stages with custom OFHC braided copper straps. As with the 300\,K shell, both the 40\,K and 4\,K shells interface with the inserted universal readout harness (URH) assemblies via rectangular cutouts~\citep{Moore_2022}. 
\par
 The 1\,K stage and 100\,mK stage are coupled to the still and mixing chamber stages of a Bluefors LD400 dilution refrigerator (DR)\footnote{https://bluefors.com/}, respectively. Mechanically, the two stages share similar designs. In order to transfer cooling power from the DR to the OTs, we utilize a hexagonal OFHC copper back-up structure (BUS). The BUS contains cutouts to minimize its overall mass while maintaining sufficient thermal conductance, and it is coupled to the DR stages via sets of braided, gold-plated OFHC copper straps\footnote{https://www.techapps.com/}. On each BUS, we bolt copper rods (``cold fingers") that protrude towards each OT. From there, another copper braided strap connects to a similar OT cold finger, completing the thermal path from the DR's mixing chamber to the OT. The 1\,K cold stage,  which is coupled to the DR's still stage, follows a similar structure. The two BUSes are mechanically connected by a carbon fiber truss. The 1\,K BUS is mounted to the 4\,K filter plate via six sets of carbon fiber tripods. \par
In order to preserve structural integrity while maintaining thermal isolation across stages, we use several sets of G10 tabs and carbon fiber trusses. Because of the importance of their functionality, these G10 tabs and carbon fiber trusses in the LATR were subjected to a series of weight tests and pull tests to assess their robustness. After over 20 cooldowns and the LATR's shipment to Chile, visual in-situ inspection of the G10 tabs and carbon fiber trusses revealed no damage, giving us confidence in their future performance~\citep{orlowski_inprep}. \par
\subsection{Readout and Detector Design} \label{sec:readout_design}
The fully populated LATR will field $\sim$62,000 TESs across 13 OTs. SO LAT deployment will be broken up into two phases: the first phase will install seven OTs into the LATR, and the second phase (referred to as Advanced Simons Observatory, or ASO) will install the final six OTs. Each OT will contain three universal focal plane modules (UFMs) of identical frequency bands.The UFMs are distributied across the following frequency bands: low-frequency (LF, with center frequencies of 27 and 39\,GHz), mid-frequency (MF, with center frequencies of 93 and 145\,GHz) and ultra-high-frequency (UHF, with center frequencies of 225 and 280\,GHz). The MF and UHF UFMs package $1756$ detectors per array (with $36$ of those being dark detectors,~\citep{Duff_24}). At time of writing, the LF UFM is still being developed and is expected to package between 222 and 244 detectors per wafer. To read out this large volume of TESs, SO employs microwave multiplexing ($\mu$Mux, \citep{McCarrick_2021}) where each TES is coupled to a resonator on a transmission line via a RF superconducting quantum interference device (SQUID). This architecture allows for the ability to install $10^3$ resonators on a single transmission line (Figure \ref{fig:mux}). MF UFM performance is reported on in detail in \cite{Dutcher_2023}. \par
\subsubsection{Cold Readout}\label{sec:cold_readout}
\begin{figure}
    \centering
    \includegraphics[width=0.7\linewidth]{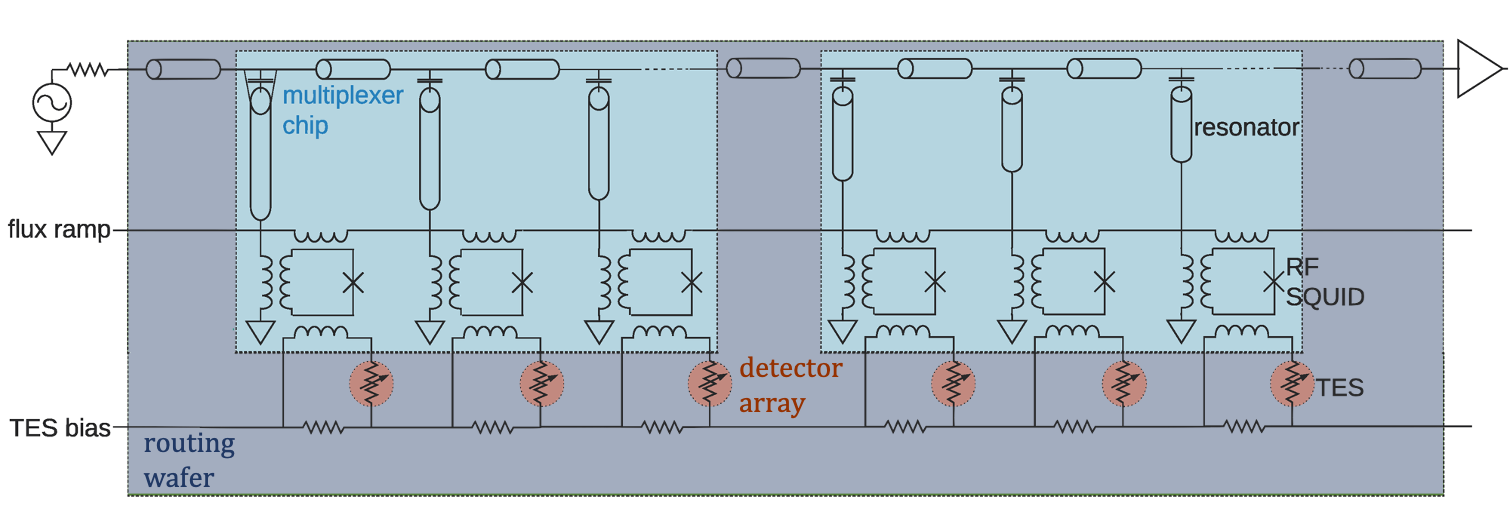}
    \caption{A diagram of the multiplexing readout design~\citep{McCarrick_2021}. A single RF transmission line is coupled to $\sim$900 resonators. Each resonator is inductively coupled to a RF SQUID, which is inductively coupled to a TES circuit. All the components in this figure are found in the ``UFM" block in Figure \ref{fig:rf_chain}. }
    \label{fig:mux}
\end{figure}
\begin{figure}
    \centering
    \includegraphics[width=.9\linewidth]{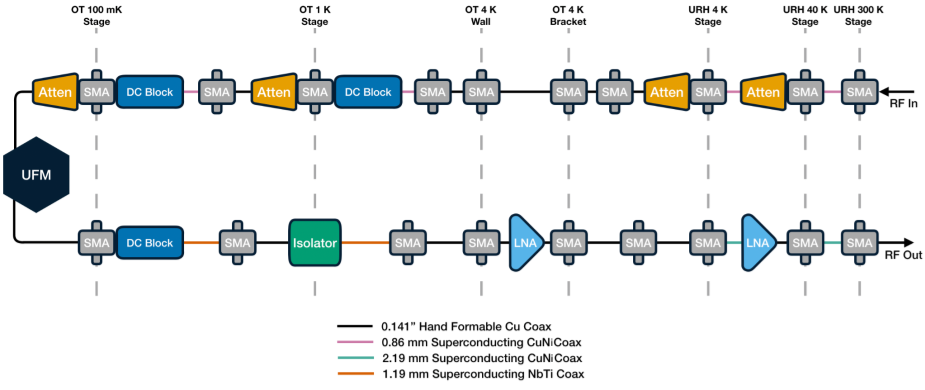}
    \caption{A diagram of the cold readout chain in the LATR. Each UFM is read out through two of these chains. On the right side of the diagram, ``RF In" and ``RF Out" are connected to the warm SMuRF electronics.}
    \label{fig:rf_chain}
\end{figure}

A single UFM contains $\sim$1800 resonators, and is read out with two sets of transmission lines, with $\sim$900 resonators coupled to each line (Figure \ref{fig:mux}). Resonator frequencies span from 4 to 6\,GHz. Additionally, each UFM has a single low frequency DC connection, used to transmit voltage biases to the TESs and flux ramping signals for the SQUIDs. Each OT contains three UFMs, and must support all the cabling and connectors necessary to read them out. To avoid strongly thermally coupling the different OT cold stages, the OT employs a combination of copper-nickel (CuNi) semi-rigid coaxial lines for input lines and niobium titanium (NbTi) coaxial lines for output lines. The CuNi coaxial lines provide attenuation for incoming signals, while the NbTi lines maintain a high signal to noise ratio for the outgoing tones.  For the same reason, the DC bias ribbon cables are made from low thermal conductivity NbTi wire. \par
Outside the OTs, readout lines are connected to the URH at 4\,K via isothermal, meter-long copper coaxial cables and DC ribbon cables. The URH is a multi-stage readout insert that carries RF transmissions, flux ramp signals, and DC bias voltages between the 4\,K stage and the 300\,K stage. Within the URH we utilize CuNi coaxial lines, and PhBr DC ribbon cables which terminate at a routing board on the 40\,K stage. Each routing board splits off bias signals to 40\,K low-noise amplifiers (LNAs) on the output RF lines. The remaining low frequency signals (i.e., flux ramps, detector biases, and 4\,K LNA biases) are routed down to the 4\,K stage via additional PhBr wiring and connectors. An analogous routing board is installed on the back of the OT at 4\,K, with a corresponding 4\,K LNA. While the LNAs are required on the output lines, cryogenic attenuators are required on the input lines to reduce the amount of power dropped on the resonators. The coaxial connections terminate with SMA connectors at 300\,K~\citep{Moore_2022, SathyanarayanaRao2020}. The full cold RF readout chain is diagrammed in Figure \ref{fig:rf_chain}. \par
\subsubsection{Warm Readout} \label{sec:warm_readout}
SO utilizes SLAC Microresonator Radio Frequency (SMuRF) electronics (\cite{Yu_2023}, \cite{Henderson_2018}) to read out LATR detectors. Each MF and UHF UFM is read out through a single SMuRF system, with six SMuRF systems populating an Advanced Telecommunications Computing Architecture (ATCA) electronics crate\footnote{https://comtel-online.com/}. Because LF detectors require larger on-array coupling optics, LF UFMs contain fewer detectors than their MF or UHF counterparts. For this reason, three LF UFMs are read out via a single SMuRF system. For seven OTs, the LATR will deploy 19 SMuRF systems across four crates. \par
A single SMuRF system consists of five integrated boards: one carrier card, one Rear Transition Module (RTM) card, two Advanced Mezzanine Cards (AMC), and one cryostat card (or ``cryocard"). Two AMCs are seated within one carrier card, with an RTM attached to the rear of the carrier card. All of these components are installed in a single ATCA crate slot. The RTM is responsible for generating the low frequency signals: the TES bias voltage, the flux ramp sawtooth signals, and the cryogenic LNA bias voltages. The AMCs generate the RF tones that are sent to the resonators through the coaxial lines, from 4 to 6\,GHz.  The cryocard relays the low frequency signals from the RTM to the cryostat via the URH. Finally, there is a room temperature LNA installed on the 300\,K side of the output lines of the URH, for further amplification. \par
The cryocard is housed in a separate RF-protected enclosure, which is mounted directly above the URH on the LATR's 300\,K shell. The SMuRF crates are mounted on racks directly on the ribs of the LATR's vacuum shell. This mechanical arrangement minimizes the length of coaxial lines between the URH 300\,K plate and the AMC, reducing total signal loss and ambient coupling~\citep{silva_2022}. \par

\subsection{Optical Design}
Light reflected off the LAT's secondary mirror enters the LATR through thirteen 3.175\,mm thick anti-reflection (AR) coated  ultra-high molecular weight polyethylene windows mounted to the 300\,K front plate. Mounted behind the windows on the 300\,K front plate is a double-sided infrared blocking filter (DSIR). The DSIRs are installed to reduce optical loading on the colder stages~\citep{ade_2006}. \par
%For the purposes of in-lab testing, the LATR employed 1/4" thick windows. However, for the deployment grade, AR coated windows, we use 1/8" thick windows. The thinner window can be used in deployment without risk of failure due to lower atmospheric pressures at the site.
On the 80\,K filter plate, we install another DSIR and an AR-coated alumina filter. The alumina filters are designed to act as prisms refracting incoming light into the OTs, allowing the OTs to be co-linear with the entire LATR. The central OT has a flat alumina filter, while the inner and outer rings of OTs have increasingly larger sloped gradients, sloping towards the cylindrical axis of the instrument. For reference, the focal plane of the two-mirror crossed-Dragone system is curved, and the edge rays are thus not parallel with the central axis of the OTs~\citep{dicker_18}.  The prism shape of the alumina filters here enables a flat detector focal plane. Between the DSIR and the alumina filter, the 80\,K stage absorbs a large quantity of optical loading (see Section \ref{sec:cryo_validation}). At 40\,K, there is one additional DSIR. \par
\begin{figure}[h]
    \centering
    \includegraphics[width=.8\textwidth]{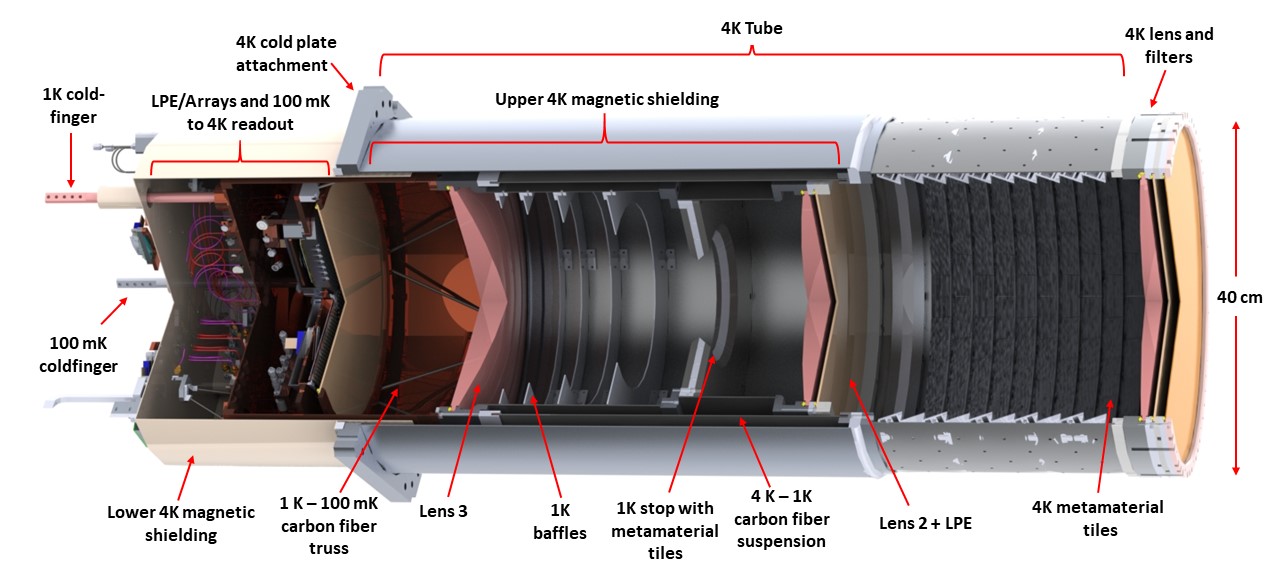}
    \caption{A rendered, cross-sectional view of the full OT \citep{Zhu_2021}. Photons enter the OT on the right side of this figure and travel leftward until they reach the detector arrays at 100\,mK (or get absorbed or reflected by another optical element first). The associated readout hardware for each array is housed in the 4\,K magnetic shielding (left). The only structural mounting point for the OTs in the LATR is the 4\,K flange (middle-left). }
    \label{fig:OT_diagram}
\end{figure}

 % The shape of the alumina filter depends on its location on the LATR: because the wedged alumina filters act as prisms for the non-central, off-axis OTs, the outer most alumina wedges have sloped gradients, while the center one is flat.
The remaining optical elements are mechanically integrated into the OTs (Figure \ref{fig:OT_diagram}). At 4\,K, there is a thick IR blocker, a low-pass edge (LPE) filter, and an AR coated silicon lens.  The LPE helps reduce optical loading from out-of-band sources. The silicon lenses are designed to re-image light from the LAT's focal plane onto the OTs' focal plane baseplates (FPB).  At 1\,K, there is a second lens, a Lyot stop and series of baffles (covered in injection molded, carbon-loaded plastic tiles \citep{Xu_2021}), and a final, third lens. At 100\,mK we install an LPE, and finally the UFMs. The UFMs contain gold-plated aluminum feedhorns, directly mounted in front of ortho-mode transducers (OMTs) that couple radiation to the TESs. The original cold optical design~\citep{dicker_18,Gudmundsson_2021} included another LPE at 1\,K, but MF OT holography measurements revealed an excess of side-lobe power sourced by this LPE. Follow-up measurements of this OT without the LPE concluded that optical loading and bandpass specifications could still be satisfied without the filter, leading to the decision to leave it out in the final deployment configuration~\citep{Sierra_2023}. \par

\subsection{LATR Performance Specifications}\label{sec:latr_specs}

%\begin{table}[h]
%\centering
%\begin{tabular}{ |c |c c c c c |}
%\hline
% \textbf{Stage} & 80\,K & 40\,K & 4\,K & 1\,K & 100\,mK \\ \hline
% \textbf{Requirement} & 120\,K & 55\,K & 8\,K & 1\,K & 100\,mK\\ \hline 
%\end{tabular}
%\caption{Cryogenic temperature specifications for each stage of the LATR. The second row represents the maximum acceptable temperature of the warmest part of the stage. We assume that this occurs at the center of optical elements, where the path for thermal conductivity is greatest.}
%\label{tab:cryo_specs}
%\end{table}

LATR design decisions were motivated by the requirement to cool the UFMs down to 100\,mK. All thermal specifications for the individual cold stages were driven by this requirement. For the warmer stages, the largest concern was about the effect of thermal radiation on colder stages, particularly within the cold optical filter stack. In Tables \ref{table:80k_loading} and \ref{table:40k_loading}, we report temperature requirements for the various LATR stages. Thermal simulations suggested that higher temperatures at 80\,K would have a minimal effect on the 100\,mK stage, so its associated maximum acceptable temperature is quite high~\citep{Hill_2018}. The reported 4\,K stage requirement describes the temperature of the OT's 4\,K filter (right-most optical element in Figure \ref{fig:OT_diagram}). While these temperature stages are not expected to directly affect detector performance, we were concerned that they could potentially warm the 100\,mK stage through radiative power. The pulse tube load curves are fairly flat in the regime close to their nominal temperatures listed in Tables \ref{table:80k_loading} and \ref{table:40k_loading}. Moreover, the conductivity of aluminum increases with temperature between 40 and 4\,K \citep{Woodcraft2009}, further alleviating the effect of increased load. Therefore, we only require that the 80, 40, and 4\,K loads be under the expected PT capacity to achieve specification. Specifications for 1\,K are driven by cooling loop performance in the DR unit. We budgeted $2.85\times10^{-3}$\,W for the 1\,K stage and $66.5\times10^{-6}$\,W for the 100\,mK stage~\citep{Zhu_2021}.  \par

%%%%%%%%%%%%%%%%%%%%%%%%%%%%%%%%%%%%%%%%%%%%%%%%%%%%%%%%%%%%%%%%%
%%%%%%%%%%%%%%%%%%%%%%%%%%%%%%%%%%%%%%%%%%%%%%%%%%%%%%%%%%%%%%%%%
%\section{LATR lab validation} \label{sec:LATR lab validation}
%%%%%%%%%%%%%%%%%%%%%%%%%%%%%%%%%%%%%%%%%%%%%%%%%%%%%%%%%%%%%%%%%
\section{Cryogenic Validation} \label{sec:cryo_validation}
%\tb{Tanay/Jack}
The LATR has previously been shown to meet cryogenic design specifications with a single, dark OT installed (\cite{Zhu_2021}). In this configuration, no OT lenses, filters, or UFMs were installed. The addition of these elements introduces new sources of thermal loading that require characterization. Through incremental testing, we find that the LATR will still meet cryogenic requirements with 13 fully assembled OTs installed, as listed in Tables \ref{table:40k_loading} and \ref{table:80k_loading}.\par
As mentioned in Section \ref{sec:Cryomechanical_Design}, any time spent cooling the LATR on the telescope is time taken away from science operations, implying that faster cooldowns can lead to deeper maps. Before turning on any pulse tubes, we pump the LATR out to $10^{-3}$\,mbar. In North America, the LATR typically pumps down to the base level pressure within four days, depending on humidity. At the site, where atmospheric pressure is roughly half of what it is at sea level, it takes just under three days to pump out. Once fully pumped out, the cryostat reaches a temperature of 4\,K within four days. After reaching 4\,K, we start condensing the $^3$He\textendash$^4$He mixture in the DR, which brings the 100\,mK stage down to its base temperature within hours. In sum, once completely assembled at the Chilean site, the LATR (without any OTs) can reach base temperature within a week. Based on incremental testing (by adding more OTs during in-lab testing), we estimate that each OT adds $\sim0.5$\,days to the cooldown time. Therefore, we project that the fully assembled and evacuated LATR (with 13 OTs) will take roughly 10 \textendash\,11 days to reach base temperature. All temperatures reported in this paper are recorded via a custom, open-source data acquisition framework~\citep{Koopman_2020}. \par
\subsection{100 mK}
\cite{Zhu_2021} showed that a single dark OT, without optical elements and UFMs installed, added $<4\,\mu$W of loading at 100\,mK. Additionally, we find that the 100\,mK thermal BUS adds $20\,\mu$W of power. We subsequently performed several cooldowns in various configurations, integrating mechanical, optical, and electronic components as they became available. Ultimately, we were able to assemble an entire OT (with all its respective optical and electronic components, and unbiased detectors) and found the total load of a single OT to be $\leq6\,\mu$W, with the FPB temperature to be 54\,mK. With this loading, we estimated the FPB temperature in a 13 OT cooldown to be $\geq91$\,mK.  \par

%\cite{Zhu_2021} showed that a single dark OT, without optical elements and UFMs installed, added $<4\mu$W of loading at 100\,mK. We subsequently performed a 7 OT dark cryogenic cooldown, which showed that the total load from mechanical components at 100\,mK to be $50\mu$W (thus $7\mu$W per OT). In this configuration, the DR ran at 44\,mK, while the hottest UFM was at 75\,mK. Of the total $\sim\Delta$30\,mK gradient between the DR and hottest UFM, $\sim\Delta$25\,mK was due to the DR strap, and the remaining $\sim\Delta$5\,mK is seen across the BUS and OT straps to the OT FPB. Since the hottest UFM is $<100$\,mK, this level of thermal loading is acceptable for a 7 OT configuration. We then extrapolated this loading to 13 OTs, resulting in an estimate of 102$\mu$W, assuming the pessimistic $7\mu$W loading value. We converted this into an estimate of the hottest UFM by first using our calibrated DR load curve to get the temperature of the DR for the 13 OT loading, 52\,mK. We then assumed the thermal gradients scaled linearly with load, resulting in an estimate that the hottest UFM would be at 110\,mK. This exceeds our cryogenic specifications. Additionally, in a separate test where we introduced sub-40\,K optical elements that allow light through to the 100\,mK stage of a single OT without arrays, we found that the estimated OT load was $\leq6\mu$W, as compared to the budgeted $5\mu$W per OT, worsening our forecast for the 13 OT configuration. \par

At 100\,mK, we budgeted $5\,\mu$W of total loading per OT; measured values from the test above suggests a higher loading than expected. In order to address this, we introduced another set of heat straps between the 100\,mK BUS and the DR to improve the thermal conductivity between them, and hence reduce the gradient induced by the higher loading levels.

In our final in-lab cooldown, we installed two fully assembled, optically-coupled OTs with the extra set of heat straps, and we estimated the 100\,mK loading to be 10\,$\mu$W (on top of the $20\,\mu$W from the thermal BUS). In order to test the new heat strap set-up, we assumed that each OT would add 5\,$\mu$W, and applied 65$\,\mu$W to a single resistive heater mounted to the 100\,mK BUS to simulate a full 13 OT configuration. At this loading, the DR mixing chamber rose to 40\,mK. However, the gradient across the DR straps was only 15\,mK (compared to 25\,mK with a single heat strap). After extrapolating and assuming the worst gradient from the BUS to the OT FPB based on measurements from previous cooldowns, we estimate that the hottest FPB temperature for a 13 OT configuration would be 80\,mK and the BUS temperature would be 55\,mK. This is comfortably below our 100\,mK requirement listed in Tables~\ref{table:80k_loading} and \ref{table:40k_loading}. \par
In 2024, we integrated the LATR with six optically coupled OTs with the LAT. We present the results of cryogenic testing for this cooldown in Section \ref{sec:LAT}.
\subsection{80K/40K/4K}
In addition to tracking the loading at 100\,mK, we also estimated the loading at 80, 40, and 4\,K, and made similar extrapolations to 13 OTs. \par 
The 80\,K stage's performance is largely independent from the rest of the receiver, and thus the easiest to characterize. In Table \ref{table:80k_loading}, we show that with 7 sets of filters, the filter plate is well under 80\,K. Assuming the gradient across the 80\,K plate scales linearly with the number of filters, extrapolations to the 13\,OT configuration using this data suggest that the hottest part of the 80\,K filter plate reaches 85\,K, which is under the specification listed in Table~\ref{table:80k_loading}. \par

The 40 and 4\,K loading is shown in Table~\ref{table:40k_loading}. Since we employ a dual-stage PT-420 pulse tube here, the performances of these two stages are coupled. This leads to some degeneracy in the measured loading values and contributes to the uncertainties reported in Table~\ref{table:40k_loading}. Additionally, we installed components as they were obtained; for this reason, the configurations reported in Table~\ref{table:40k_loading} are not necessarily well-defined. For example, the 1\,OT setup includes one OT, one URH, and three filter sets at 40\,K, meaning that from the perspective of 4\,K, this is a 1\,OT cooldown, while from the perspective of 40\,K, it is a 3\,OT cooldown. The 40 and 4\,K loading is not strictly linear with number of OTs due to slightly differing numbers of available readout cables between OTs. This is a small effect, and to a good approximation, we can estimate the 13 OT loading as:

\begin{equation}
    \text{q}_{13} = \text{q}_{\text{dark}} + \frac{13}{7}(\text{q}_{7} - \text{q}_{\text{dark}}),
    \label{eq:thermal_extrapolation}
\end{equation}
where $q_{dark}$ represents loading in a zero OT configuration. This results in a loading estimate of $66\pm 2$\,W and $1.7\pm0.2$\,W at 40 and 4\,K, respectively. The 40\,K loading is somewhat higher than budgeted \citep{Zhu_2021}, while the 4\,K loading is lower than budgeted. Both are, however, significantly less than the 40 and 4\,K capacities of 110\,W and 4\,W, respectively. As such we have qualified the 40 and 4\,K stages for 7 OTs and forecast that they will also meet specifications for 13 OTs, as described in Tables \ref{table:80k_loading} and \ref{table:40k_loading}.

\par
\begin{table}[h!]
\centering
\begin{tabular}{c|| c c } 
 \hline
 Configuration & Filter Plate Temperature &  Loading  \\ [0.5ex] 
 \hline\hline
  Budgeted & $<$120\,K & 180\,W \\
 \hline
 Dark & 37-39\,K & 22$^{+1}_{-1}$\,W   \\ 
 2 Filters & 40-43\,K & 35$^{+1}_{-1}$\,W \\
 3 Filters & 44-47\,K & 42$^{+1}_{-1}$\,W \\
 7 Filters & 48-58\,K& 88$^{+1}_{-1}$\,W \\
 13 Filters* & 63-85\,K & 144$^{+1}_{-1}$\,W  \\ [1ex] 
 \hline
 6 OTs & 40-48\,K & 47$^{+1}_{-1}$\,W  \\ 
 \hline
\end{tabular}
\caption{Budget and estimates of 80\,K loading. The first row (``Budgeted") are informed by performance specfications~\citep{Zhu_2021}. The number of filters in the configuration column refer to the number of 300\,K windows, 300\,K DSIRs, and 80\,K DSIRs that are installed. Values for the asterisked configuration come from an estimate by extrapolation of measured values, using Equation \ref{eq:thermal_extrapolation}. The ``Filter Plate Temperature" describes the range of temperatures measured across the entire filter plate, where the coldest temperature is at one of the pulse tube cryocooler. The final row (``6 OTs") details the most recent cryogenic measurements from the LATR, with a different filter stack than the other rows; see Section \ref{sec:LAT} for more details. }
\label{table:80k_loading}
\end{table}
\begin{table}[h!]
\centering
\begin{tabular}{c|| c c | c c} 
 \hline
 Configuration & 4\,K Plate Temperature & 4\,K Loading & 40\,K Plate temperature & 40\,K Loading \\ [0.5ex] 
 \hline\hline
Budgeted & $<$8\,K & 4.0\,W & $<$55\,K & 110\,W \\
\hline
 Dark & 2.7 - 5.0\,K & $0.9^{+1}_{-1}$\,W &  $27-47$\,K & $33^{+1}_{-1}$\,W  \\ %2019/10/23
 %2 Filters& 3.1 - 6.2\,K & & &   \\
 1 OT & 2.9 - 5.3\,K & $1.0^{+0.1}_{-0.1}$\,W& $27-48$\,K &  $34^{+1}_{-1}$\,W \\ %2020/02/20
 3 OT & 2.9 - 5.4\,K &$1.1^{+0.1}_{-0.1}$\,W & $28-49$\,K & $38^{+1}_{-1}$\,W \\ %2021/02/16
 7 OT & 3.0 - 6.3\,K & $1.3^{+0.1}_{-0.1}$\,W& $30-53$\,K & $51^{+1}_{-1}$\,W \\ %2021/08/20 
13 OT* & $>$3.5\,K & $1.7^{+0.2}_{-0.2}$\,W& $>$31\,K & $66^{+2}_{-2}$\,W \\[1ex] 
 \hline
 6 OTs & 3.0 - 6.3\,K & $0.7^{+0.2}_{-0.2}$\,W& 32 - 52K & $61^{+1}_{-1}$\,W \\
 %13 OTs* & 3.0 - 6.3\,K & $0.7^{+0.2}_{-0.2}$\,W& 32 - 52K & $93^{+2}_{-2}$\,W \\[1ex] 
 \hline
\end{tabular}
\caption{Budget and estimates of 4\.K and 40\,K Loading. The dark setup features no OTs, no URHs, and no filter sets. The first row (``Budgeted") are informed by performance specfications~\citep{Zhu_2021}. The ``1 OT" setup includes one OT, one URH, and three filter sets. The ``3OT" setup includes three OTs, one URH, and two filter sets. The ``7OT" setup includes seven OTs, two URHs, and seven filter sets. The values in the final row (``13OT") represent estimates, from extrapolations based on measurements from the first several rows of this table. The "Plate Temperature" column describes the range of temperatures measured across the entire filter plate, where the coldest temperature is at one of the pulse tube cryocooler. The final row (``6 OTs") details the most recent cryogenic measurements from the LATR, with a different filter stack than the other rows; see Section \ref{sec:LAT} for more details.}
\label{table:40k_loading}
\end{table}
\subsection{Thermal Stability} \label{sec:thermal_stability}
\begin{figure}[h]
    \centering
    \includegraphics[scale=0.6]{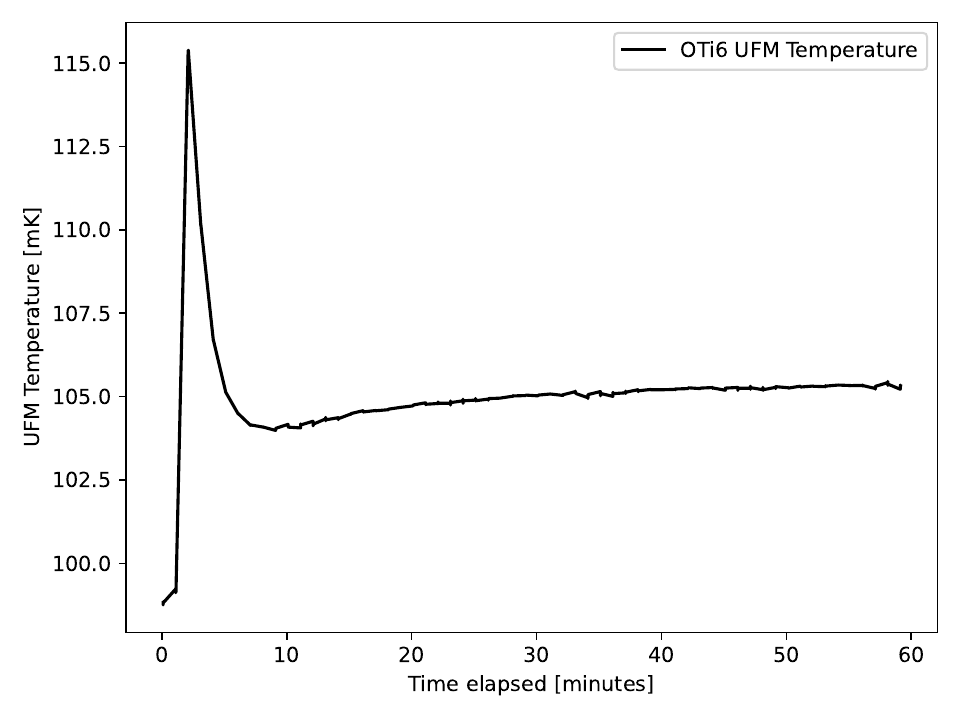}
    \caption{The temperature of an UFM as it is being overbiased and dropped into the transition phase during in-lab testing. The overbias results in a temperature change of $\Delta15$\,mK from the initial $T_\mathrm{bath}$ and then drops to operating temperature, where it remains stable. Note that the UFM is hotter than temperatures presented in Section \ref{sec:cryo_validation}; in order to bias detectors easily, we heated the DR mixing chamber stage to put the UFM temperatures at exactly 100\,mK. }
    \label{fig:UFM_Temperature_Stability}
\end{figure}

Thermal fluctuations on the UFMs can lead to high 1/$f$ noise levels and, in some cases, a loss of resonator tracking. In order to avoid this, the LATR's 100\,mK stage must remain as thermally stable as possible. In a static state, the OT FPBs have been shown to be stable to millikelvin levels~\citep{Bhandarkar_2022}. However, when overbiasing detectors, the loading at the 100\,mK stage jumps by 500\,nW per array~\citep{wang_2022, Dutcher_2023}. This is a sudden increase in power on the cold stage, but as seen in Figure \ref{fig:UFM_Temperature_Stability}, the cold stage recovers to a stable operating temperature in 5 minutes. When UFMs are biased into their transitions, the load falls to 150\,nW per UFM. Detector specifications require thermal fluctuations on the UFMs to be $\leq 500$~$\mathrm{\mu K}$ over 5 minutes. After allowing detectors to settle following biasing, we find the root-mean-square of the UFM temperature across any five minute time period to be $\sim310$~$\mathrm{\mu K}$. Here, we take the 12 minute mark in Figure \ref{fig:UFM_Temperature_Stability} to indicate the start of the stable period. \par
\subsection{TES response vs Bath Temperature} \label{sec:bath_temp}

TES bolometers are sensitive to changes in the temperature of the $100$\,mK stage of the cryostat to which they sink power ($T_\mathrm{bath}$).  To maintain equilibrium, the power coming into the TES, including the photons it measures and electrical heating from the TES bias, must match the power flowing out to the bath, which depends on the temperature difference between the TES and $T_\mathrm{bath}$.  To first order, small variations in $T_\mathrm{bath}$ look like signal in the detector timestreams.

The coupling factor -- measured in $\mathrm{pA/mK}$ or $\mathrm{pW/mK}$ -- between changes in $T_{bath}$ and detector signal can be calculated from known detector properties, including the thermal conductivity between the TES and the bath and TES gain (which depends on bias level (\cite{irwin_hilton_2005})). %\ji{check that this is where I got info from when doing this calculation before, or at least includes the relevant info}.

Other parts of the detector and readout system are also temperature sensitive, such as the resonators in the readout chain or the impedance of the (non-superconducting) coax.  The TESs will also slightly vary in gain (current response to incoming power) with changes in $T_\mathrm{bath}$.  However, the dominant response to changes in $T_\mathrm{bath}$ should be the apparent signal picked up in the TESs. Understanding this coupling also has implications for stability requirements on $T_\mathrm{bath}$ while operating and for data processing (see Section \ref{sec:thermal_stability}).

\begin{figure}
    \centering
    \includegraphics[scale=0.5]{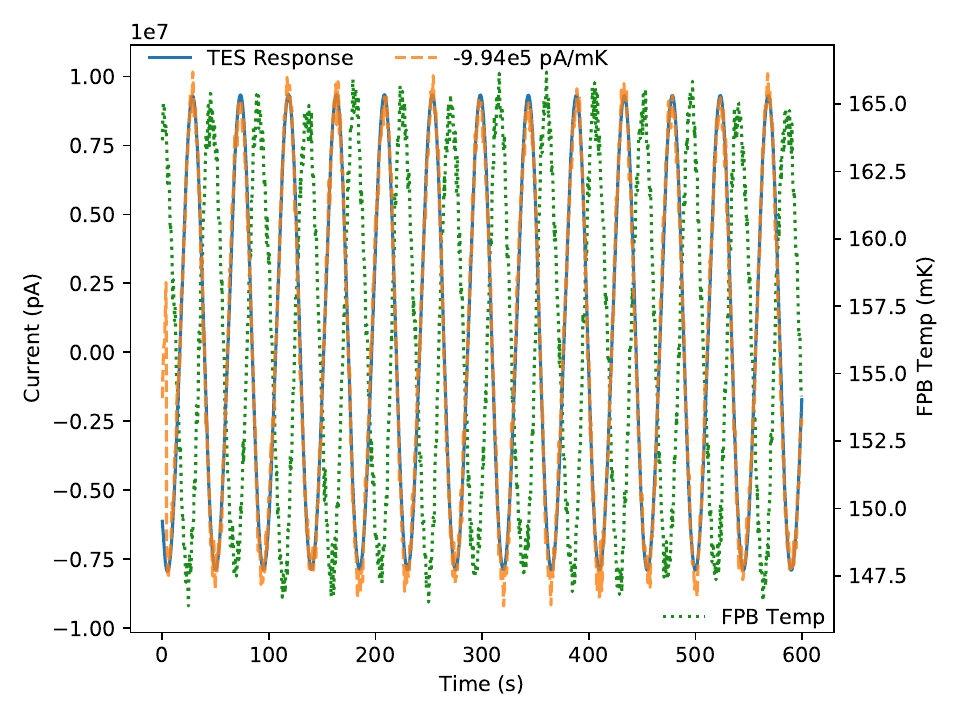}
    \includegraphics[scale=0.5]{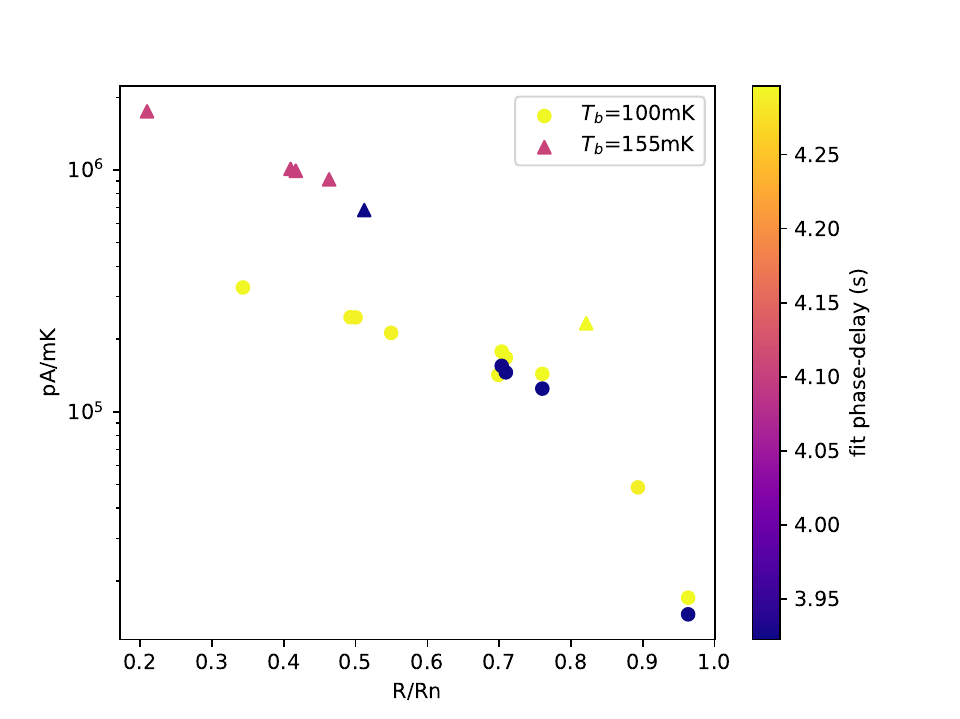}
    \caption{\emph{Left:} Example TES response (blue, left-axis) compared to oscillating FPB temperature (dotted green, right-axis). The temperature profile is fit with with a sine wave for an amplitude (the coupling factor) and phase-delay.  The 4\,s delay between the recorded temperature and TES response results from the thermometer being  closer to the heater than the TES. \emph{Right:} A summary of measured coupling factors compared to fraction of normal resistance each TES was biased to, and separating the tests at different bath temperatures. Non-functioning or unbiased detectors were removed.  These results are consistent with the values predicted from the detector properties.  This demonstrates that the readout system is not excessively sensitive to thermal changes, and so the known TES response dominates 100\,mK thermal sensitivity in the LATR.
    }
    \label{fig:Tbath_var}
\end{figure}

To determine the coupling factor, we designed a test to vary $T_\mathrm{bath}$ by varying the power on a resisitve heater mounted on the OT's FPB. For our initial in-lab testing, we installed a high conductance single pixel box (SPB), with 6 TESs. Testing with an SPB allowed for small-scale detector and readout testing, before scaling up to deployment-grade UFMs. With a function generator, we powered the heater with a long period (45\,s) sine wave, in order to give time for the thermal mass of the FPB to respond to the change in power on the resistor. From this test, we estimated a coupling factor of order $10^6\,\mathrm{pA/mK}$, which was consistent with predictions (Figure \ref{fig:Tbath_var}). \par

However, we note that the high conductance nature of the SPB is not representative of deployment-grade UFMs. Once we had installed UFMs into the LATR, we ran similar tests to determine the coupling factor. For these tests, we incrementally increased the temperature of the UFMs while recording time-ordered data (TODs). While detectors were biased to 50\%$R_N$, we measured the coupling factor to be $0.025$\,$\mathrm{pW/mK}$, . \par
We note that there is a difference in units presented here (pW vs. pA). During SPB testing, we were primarily focused on validating readout noise. Current units are a helpful gauge for this, and much of our infrastructure was designed to test this. However, as we advanced to full-scale detector testing, power units became more important and relevant. Thus, when we tested UFMs, we took measurements in units of pW. \par %It is difficult to compare the two different measurements, because _____, but we found that both iterations of testing produced results that matched expectations. \par 

During LATR testing, we were concerned about the effect telescope motion may have on the receiver's cryogenic environment. This concern motivated the measurements and tests described in this section; because we could not directly measure the effect of LAT motion before LAT/LATR integration,  characterizing and constraining the effect of changes to $T_{bath}$ helped manage this risk. In Section \ref{sec:LAT}, we will describe the effect of LAT motions on the LATR's cryogenic and detector systems.

%The primary test performed was to vary $T_{bath}$ using a heater mounted on the OT's FPB, driven by a function generator.   We ran these tests on a high conductance single pixel box (SPB), with 6 TESs, repeated at various bias points and with $T_{bath}$ at $100$\,mK and $155$\,mK.  Given the properties of these TESs, we would predict a coupling factor of order $10^6\,pA/mK$, depending on bath temperature and bias level, which is consistent with our testing. 

%Figure \ref{fig:Tbath_var} shows an example $T_{bath}$ variation and response for one detector alongside a summary of the measured coupling factors in the SPB, demonstrating that the detectors behave as expected.  

%We note that the results derived from these SPB tests may be misleading, because the SPB was designed to have a higher thermal conductance than a UFM would have. For this reason, we carried out a series of similar tests with deployment-grade UFMs, and found the coupling factor to be 0.025\,pW/mK. In Section \ref{sec:LAT}, we describe these tests in more detail. \tb{Should I move the details from Section 8 to this section here?}

%%%%%%%%%%%%%%%%%%%%%%%%%%%%%%%%%%%%%%%%%%%%%%%%%%%%%%%%%%%%%%%%%%%%%%%%%%%
\section{Readout and Detector Validation} \label{sec:rd_validation}
Following cryogenic validation, we focused on testing the LATR's readout system. Characterizing the detector system and understanding the contributions of environmental effects to readout and detector noise prior to on-sky observations was critical in the receiver's validation efforts. For the baseline detector properties presented in this section, the OT was closed off with an aluminum plate at 4\,K. Thus, all the UFMs were exposed to 4\,K radiation. \par
For each cooldown, the LATR readout chain is subject to a litany of checks before closing up: continuity checks are conducted for each DC line, every 4\,K and 40\,K LNA is biased to check functionality, and the RF chain is looped back at the 4\,K stage on the URH. For this last check, we check the 3\,GHz to 9\,GHz coaxial transmission from 300\,K to 4\,K with a vector network analyzer (VNA). This warm transmission measurement is useful for checking cable loss and poor coaxial connections. Each UFM bias line and flux ramp line is probed for short or open circuits twice; once before installation into the OT, and once after being fully integrated in an OT. After the OT is installed in the LATR, the full readout chain is connected, from the URH's 300\,K stage to the UFMs on OT's 100\,mK FPB. In this stage of assembly, we can probe the bias lines one last time; however, we cannot make warm RF transmission measurements once the resonators are installed, because the resistance of the niobium traces on the devices is too large to make RF transmission measurements at room temperature. Finally, we connect SMuRF systems and run basic system health checks on the warm hardware. \par
Once all physical connections and SMuRF systems pass their health checks, we cool the LATR down and start the detector tuning and biasing processes. To begin, we run a full transmission sweep and identify all the resonators (this often includes estimating and setting amplifier biases and in-line attenuation values). We then start tracking resonators, which allows us to monitor TES current for changes in absorbed power. At this point, we are able to stream data and estimate noise-equivalent currents (NEI), providing us with useful information about system performance. If superconducting detector noise is consistent with expectations, we then begin biasing the TESs. This involves running I-V load curves, where we drive the detectors normal by overbiasing them and then slowly step the bias value down to zero. This brings the TESs through their transition regime and provides us with the values of desired bias voltages for operations (typically at 50\%\,$R/R_N$, where $R_N$ is the normal resistance of the TES). After setting up detectors initially, we cache all relevant parameters and refer to those values when re-tuning. Typically, these values do not change greatly during a single cooldown. Therefore, caching values cuts out some of the initial set-up steps and minimizes the time spent setting up detectors if the need to re-tune or re-bias comes up. All the processes described here are relayed to the SMuRF hardware via custom software.\footnote{https://github.com/simonsobs/sodetlib}$^{,}$\footnote{https://github.com/slaclab/pysmurf} \par
\subsection{Baseline Noise} \label{sec:baseline_noise}

Initial validation of the LATR's readout system was conducted with an SPB containing 66 RF resonators (\cite{xu_2020}). The simplicity of the SPB enabled straightforward baseline characterization of the readout chain; it allowed us to understand basic properties of the readout system (such as cable loss) while avoiding the confusion introduced from a large number of detector channels. Six of the 66 resonators in the SPB are coupled to prototype TESs, allowing us to test the DC bias chains as well. Testing of the SPB showed that the LATR's readout chain was performing as expected, in terms of RF noise and DC continuity. \par
Following SPB testing, we integrated UFMs in to the LATR and continued baseline testing. An MF UFM is designed to support 1756 TESs, but it will never have 100\% yield rates when testing and fielding these devices.  
For each UFM, SO aims to field detectors at a $\geq70\%$ end-to-end yield, with an ultimate goal of  $80\%$ \citep{McCarrick_2021, Dutcher_2023}. In Table \ref{tab:det_yield} we list the measured detector yields for six UFMs installed in the LATR. Yield is determined by observing the TES responses to I-V load curves. TES that do not respond to I-V curves correctly are cut and are not a part of the reported yield counts. In Table \ref{tab:det_yield}, we show that the first six deployed UFMs meet this specification. UFM Mv11 has a noticeably lower yield; this is due to an open bias line that was discovered after LATR assembly. We know that there are 135 TES associated with this bias line, and the yield rate across the other eleven bias lines on Mv11 is $79.8\%$.  \par
\begin{figure}
    \centering
    \includegraphics[width=0.75\textwidth]{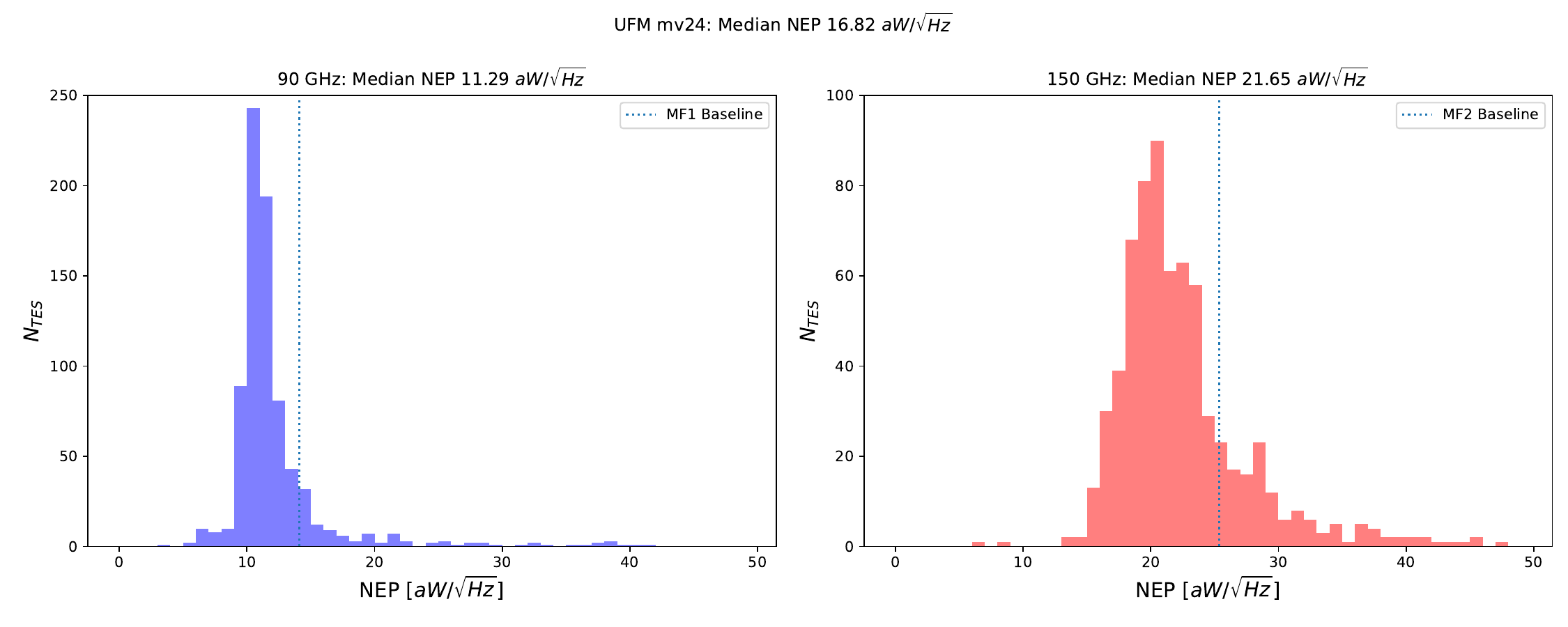}
    \caption{NEPs for a dark MF array installed in the LATR, with detectors biased on transition. The vertical dotted lines represent dark baseline specifications set for the LATR's MF UFMs (14.12\,$\mathrm{aW/\sqrt{Hz}}$ for 90\,GHz, and 25.36\,$\mathrm{aW/\sqrt{Hz}}$ for 150\,GHz). We can split the TES into 90\,GHz and 150\,GHz populations based on associated bias lines.}
    \label{fig:NEPS_by_BG}
\end{figure}

\begin{table}[h]
    \centering
    \begin{tabular}{r||c | c| c}
        \hline
        Array & Test Yield & LATR Yield & Yield Rate\\
        \hline
        \hline
         Mv11 & 1275 & 1265 & 72.0\%\\
         Mv21 & 1381 & 1459 & 83.1\%\\
         Mv24 & 1497 & 1543 & 87.9\%\\
         Mv25 & 1582 & 1592 & 90.7\%\\
         Mv26 & 1567 & 1580 & 90.0\%\\
         Mv28 & 1442 & 1440 & 82.0\%\\
         \hline
    \end{tabular}
    \caption{Detector yield for LATR MF UFMs during dark testing. Test yield indicates yield from the pre-deployment screening in a test cryostat~\citep{Dutcher_2023}. LATR yield represents the TES yield measured in the LATR while in the field. Note that the yield for Mv11 is much lower than the other UFMs due to an open bias line, which we plan to repair. Yield rate is calculated for the LATR Yield with respect to the total number of TES per UFM, 1756.}
    \label{tab:det_yield}
\end{table}
Because LATR cooldowns are lengthy, full-scale initial in-lab UFM screening and testing happens in a test cryostat before installation in the LATR (\cite{wang_2022}). This process delivers baseline noise and efficiency measurements that can be compared with the LATR's in-situ measurements for validation purposes. Noise-equivalent power (NEP) measurements serve as our principal check for LATR UFM noise properties. Through the software mentioned in Section \ref{sec:rd_validation}, we directly measure NEI for devices under test. NEP is then calculated as the following (\cite{irwin_hilton_2005}):
\begin{equation}
    NEP = NEI \frac{1}{|s_i|} ,
\end{equation}
where $s_i$ is the responsivity of the TES, determined through bias-steps or I-Vs (\cite{wang_2022}). In Figure \ref{fig:NEPS_by_BG}, we plot the NEPs for a single biased MF UFM and show that the majority of TESs meet the dark baseline specification set for SO, derived from SO instrument forecasting models~\citep{McCarrick_2021, so_forecast_2019}. The vertical dotted lines in Figure \ref{fig:NEPS_by_BG} represent the baseline dark specifications for each MF frequency: 14.12\,$\mathrm{aW/\sqrt{Hz}}$ for 90\,GHz, and 25.36\,$\mathrm{aW/\sqrt{Hz}}$ for 150\,GHz. Future studies will be able to comment on LAT on-sky detector noise, as well as LF and UHF arrays. \par

\pagebreak
\section{In Lab Environmental Testing} \label{sec:in_lab}
\subsection{Vibrations}

\begin{figure}
    \centering
    \includegraphics[scale=0.75]{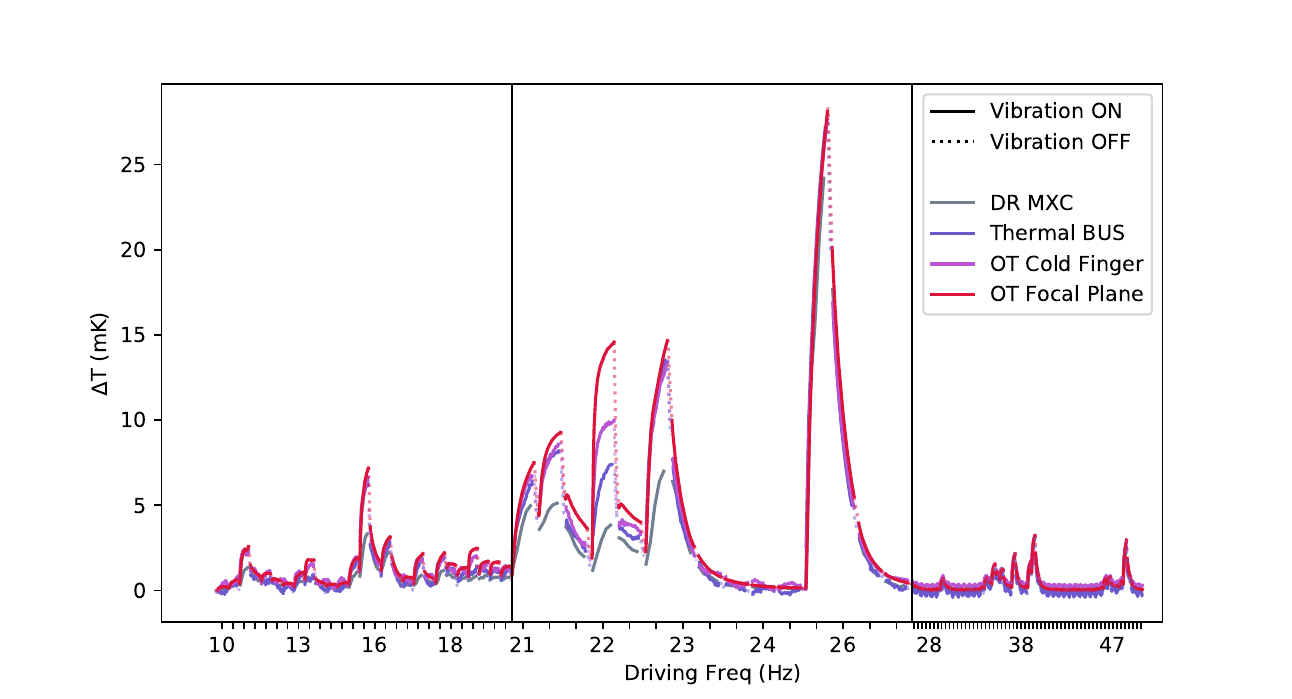}
    \caption{The thermal response to vibrations at the 100\,mK stage.  Change in thermometer temperature over time is shown, as the vibration source sweeps through discrete frequencies (with breaks in between, where the temperature is shown in dashed lines).  The x-axis has ticks and labels corresponding to each driving frequency. The largest response in this system is seen at a driving frequency of  $\sim$26\,Hz; in the LAT, the largest resonance we measured occurred at 3\,Hz, and we found no significant resonant response between 21\textendash28\,Hz. }
    
    %The largest response is seen at the OT focal planes \ji{ wasn't it usually at the cold fingers?} \tb{not according to this plot}, at $\sim$26\,Hz.}
    \label{fig:latr_vibrations}
\end{figure}

To characterize the LATR response to vibrations, we ran a series of shake tests to determine vibrational pickup within the LATR. We were initially concerned that vibrational pickup would have the largest impact in the form of readout noise, but we discovered that the larger effect was thermal. In fact, any effect seen in the detector system is a response to thermal changes induced by vibrations. We mounted a ButtKicker Mini Concert transducer\footnote{https://thebuttkicker.com/} to the flange of the LATR's front plate, and used it to produce vibrations at frequencies supplied via a tone generator. Testing shows that the most noticeable amount of vibrational pickup came in at the 100\,mK stage, in the 26\,Hz range, although there are peaks at frequencies from 21\textendash24\,Hz as well (Figure \ref{fig:latr_vibrations}). \par 
Based on simulations and our benchtop thermal BUS testing, we determined that the most likely mode of vibrations was a drum-like feature, where the largest amplitude of vibrations occur at the center of the BUS. To mitigate this, we bolted a copper bar across one axis of the hexagonal BUS, hoping to stiffen any drum-like mode. Testing showed that we did not completely eliminate the mode, but still reduced the amplitude of the thermal change. \par
As a part of factory acceptance testing of the LAT prior to its installation in Chile, we measured the vibrational environment of the receiver cabin during test scans. To do so, we installed an accelorometer to the front flange of a LATR mass dummy in the telescope and performed scans. Testing showed that the LAT's primary resonant frequency while scanning was at 3\,Hz. This finding was in agreement with pre-fabrication design studies. At higher frequencies, we found minimal power while scanning. Since most of our in-lab LATR tests raised concerns for resonant frequencies above 20\,Hz (Figure \ref{fig:latr_vibrations}), we concluded that the LATR would not experience vibrational heating when installed in the LAT. To alleviate concerns about the 3\,Hz mode, we manually excited a 3\,Hz mode on the LATR with a hammer in the lab, and saw no resultant heating. It is worth noting that any testing we conducted involved inducing more vibrational power on the LATR than it would ever actually experience in the LAT. For this reason, we are confident that any vibrational coupling in the telescope would not result in the large thermal changes shown in Figure \ref{fig:latr_vibrations}. In Section~\ref{sec:LAT}, we show that the LATR does not experience significant heating during LAT scans due to vibrations in the system. \par

\subsection{Ambient RF}

\section{LATR and LAT Integration} \label{sec:LAT}
\begin{figure}[h]
    \centering
    \includegraphics[width=0.75\textwidth]{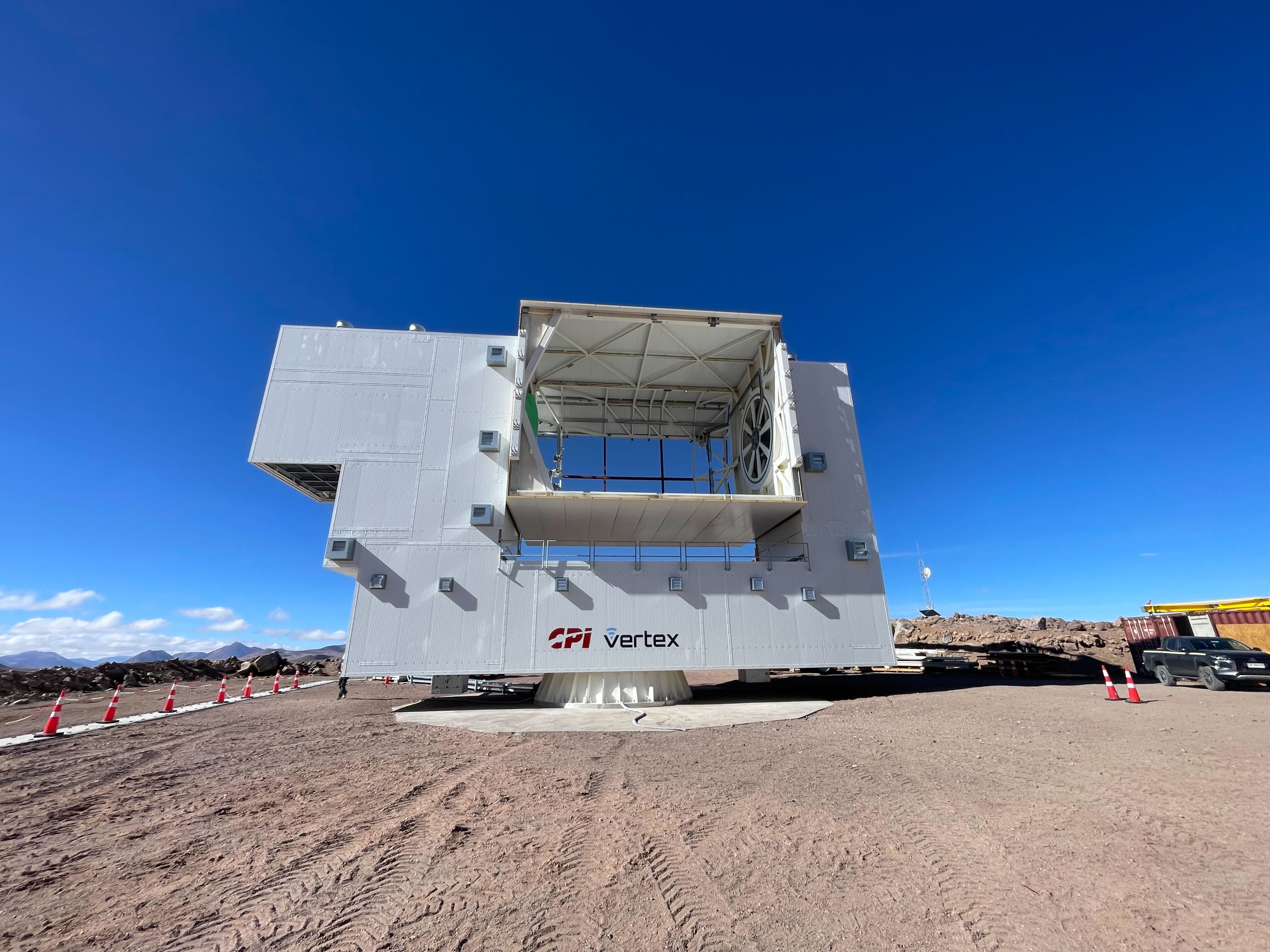}
    \caption{A photo of the LAT at the Chilean site, as of March 2024. In this image, the receiver cabin is located towards the top left of the structure. }
    \label{fig:lat_photo}
\end{figure}
Though the LAT structure has been erected at its site in Chile (Figure \ref{fig:lat_photo}), as of August 2024 the primary and secondary mirrors have yet to be fully assembled and integrated.
In order to proceed with LAT characterization, we opted to install the LATR (with two MF OTs) in the LAT, with aluminum blanks installed at the 4\,K stage of the OTs in August 2023.
Without the blank and mirrors, the TESs would be saturated and thus unusable for testing.
However, with the blanks installed, we are able to bias the TESs into transition and take dark scans while monitoring detector performance. \par
\begin{figure}[h]
    \centering
    \subfloat[Before common-mode subtraction]{
        \includegraphics[width=.5\textwidth]{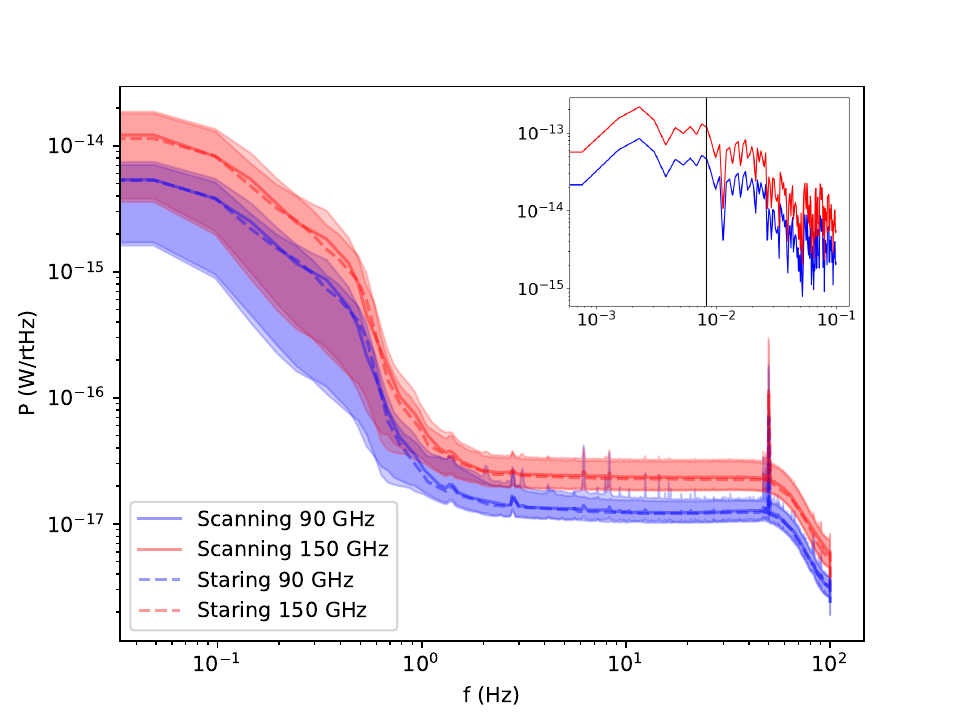}\label{fig:ASD}
       }
    \subfloat[After common-mode subtraction]{
        \includegraphics[width=.45\textwidth]{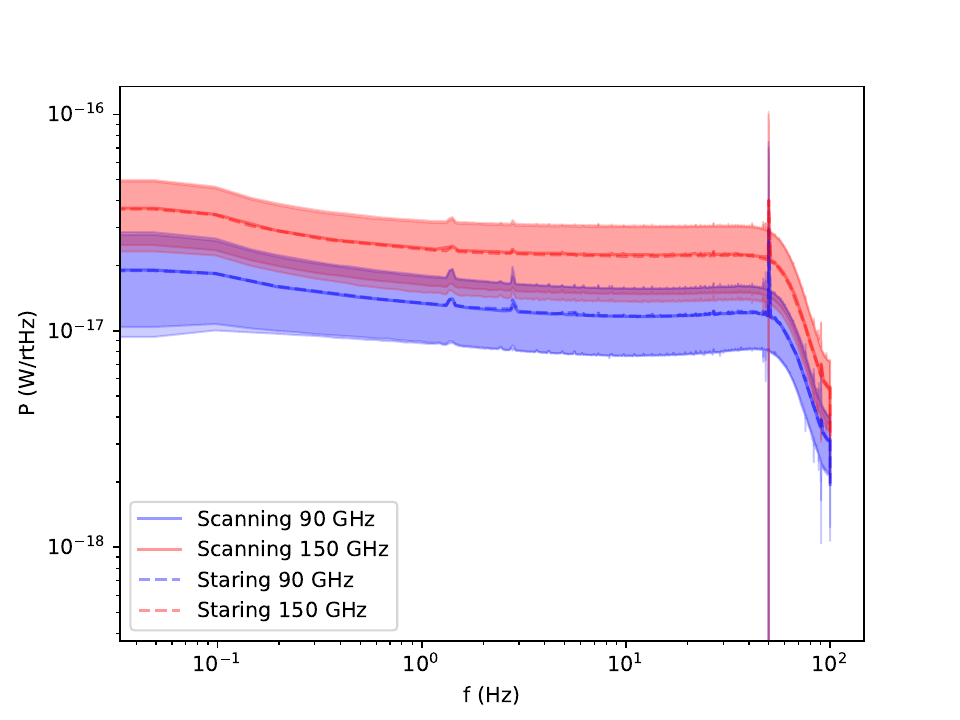}\label{fig:ASD_cmsub}
    }
    \caption{The auto-spectral density (ASD) of all detectors before and after common-mode subtraction. The solid line is the median of all detectors when scanning and the dashed line is the median of all detectors when staring. With the contour being the 1 $\sigma$ level of these distributions. Blue is the 90 GHz detectors and red is the 150 GHz detectors. Note that the solid and dashed lines sit on top or very close to each other across the entire frequency range. The spikes in the ASD before removing the common mode are mostly sourced from an issue with a single UFM. The remaining bumps after common-mode subtraction are 50 Hz (AC power), 1.4 Hz (pulse tube operating frequency), and 2.8 Hz (harmonic of the 1.4 Hz bump) and coupled to the detectors in complex ways that are not common mode. The inset plot on Figure \ref{fig:ASD} shows the ASD in the frequency range of our turnarounds, with the black vertical line being the turnaround frequency (0.008\,Hz). Figure from \cite{Haridas_2024}}.
\end{figure}

Figure \ref{fig:ASD} shows noise performance of LATR detectors during a constant elevation scan (solid lines), with the LAT moving at 1 deg/s with a throw of $60^\circ$ and an average turnaround acceleration $1$ deg/s$^2$ for $\sim 1$ week, as well as noise performance with the telescope stationary (dashed lines).
These scans mimic the expected scan strategy for LAT science observation,
so the high degree of agreement between the noise while scanning and while stationary, with very little disagreement between the two sets of auto-spectral densities, indicates  the instrument will be stable during operations.
Scans are organized, written and commanded via custom observatory control software~\citep{koopman2024}.
The peak at 50\,Hz is a result of AC power, and the peaks at 1.4\,Hz and 2.8\,Hz are caused by the pulse tube cryocoolers used by the LATR.
The low frequency noise is mostly common mode signal, and $\sim99\%$ can be removed by subtracting off a common mode modeled using singular value decomposition (SVD) (Figure \ref{fig:ASD_cmsub}).
This is a promising sign for future on-sky observations, as the LAT plans to perform maximum likelihood map-making with a noise modeling scheme similar to that of ACT, which uses SVD derived modes to model the noise \citep{Dunner_2012}. \par
\begin{figure}[h]
    \centering
    \includegraphics[scale=.65]{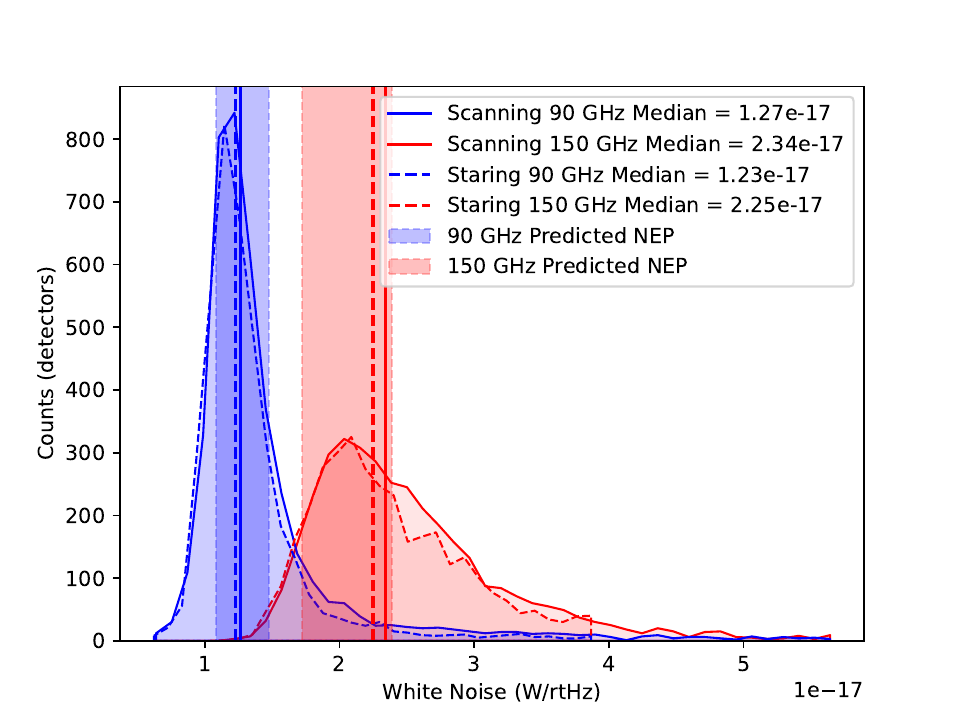}   
    \caption{White noise levels of our TODs with OTs capped at 4K. The solid line is when scanning and the dashed is when staring. Blue is the 90 GHz detectors, and red is the 150 GHz detectors. The predicted bands are the expected noise for the dark configuration from BoloCalc. Figure from \cite{Haridas_2024}
}
    \label{fig:wn}
\end{figure}

As seen in Figure \ref{fig:wn}, most detectors have a white noise level below what we predict from forecasted sensitivity simulations, and NEP medians are comparable to pre-deployment screening values \citep{Dutcher_2023}.
This implies that the derived values for LAT sensitivities used in forecasting will be achievable once the LAT is taking on-sky observations. The forecasted sensitivities were calculated using BoloCalc \citep{Hill_2018}, a sensitivity calculator that takes into account detector properties, bandpasses, the transmission and loading of all elements in the optical chain, and a model of the atmosphere at the SO site. Wherever possible, such as with the bandpasses, we use in-lab measurements for these values \citep{Sierra_2023}.\par
\begin{figure}[ht]
    \centering
    \subfloat{
        \includegraphics[width=.5\textwidth]{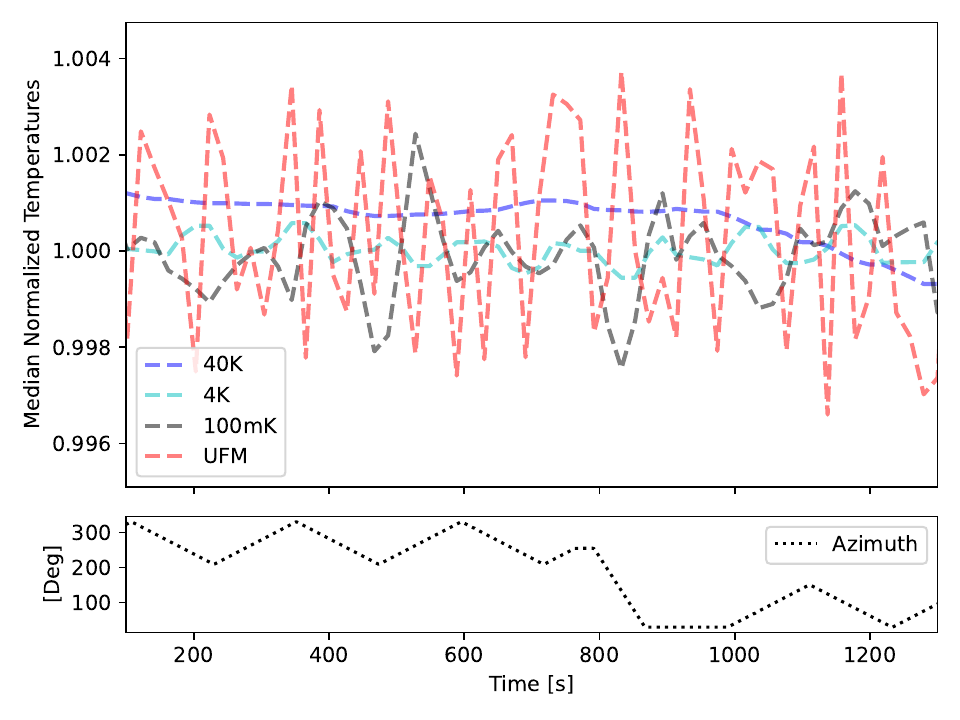}
        \label{fig:lat_temp_response}
    }
    \subfloat{
        \includegraphics[width=.5\textwidth]{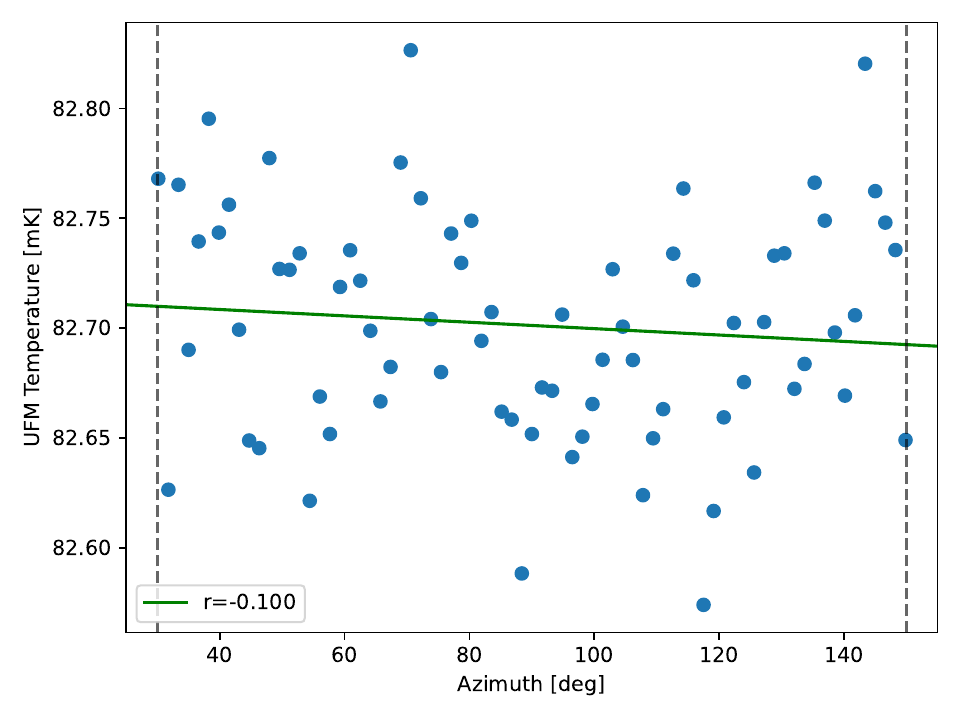}
        \label{fig:binned_az_temps}
    }
    \caption{\emph{Left: }The temperature of LATR cold stages during constant elevation LAT scans and slewing, divided by their median during this period (upper plot) and the telescope's azimuth (lower plot). We see a sub-percent level temperature change on these cold stages, which is well within specifications. Most of the noise seen in the 4\,K and 100\,mK stage can be attributed to readout noise. Slewing and scanning periods are both included because their respective parameters differ; see text for more information. \emph{Right: }Temperatures at the UFM, binned over 5 constant-elevation scans (10 total throws) from $30^\circ$ to $150^\circ$. The dotted black vertical lines indicate the azimuth angles that turnarounds occur, and the green solid line is a line of best fit, with a linear correlation coeffeceint of $r= -0.100$. We find the peak-to-peak temperature swings to be $\sim0.25$\,mK or less.}
    %\label{fig:maps}
\end{figure}

As mentioned in Section \ref{sec:in_lab}, detector performance is strongly linked to thermal stability at the 100\,mK stage.
For this reason, we were concerned about the effect LAT movement may have on cold stage temperatures, and thus, on detector performance.
In Figure \ref{fig:lat_temp_response}, we can see the effect of LAT scanning movement on the cold stages of the LATR. All temperature responses that are recorded during LAT scans are sub-percent level. We also note that at turnarounds in LAT scans, there are no significant spikes in temperature, which was a concern prior to integration. To further illustrate this, we plot the temperature of an UFM as a function of azimuth (in Figure \ref{fig:binned_az_temps}) during constant elevation scans. The total measured peak-to-peak temperature during these scans is $\sim0.25$\,mK. The turnarounds are represented by the vertical dotted-lines in this plot; we see no major temperature spikes at those azimuths, implying that the turnarounds do not disrupt the cryogenic environment of the UFM. 
Based on our test results from Section \ref{sec:bath_temp}, we found the coupling factor to be 0.025\,pW/mK.
Applying this conversion to the measured peak-to-peak temperature from above yields $6.25$x$10^{-3}$ pW as the maximum power seen by the detectors from the thermal fluctuations.
Assuming a 10\,K sky and using the baseline BoloCalc conversion for this configuration (Table \ref{table:bolocalc}), we expect sky loading will be $\sim0.43$\,pW at 90 GHz and $\sim0.59$\,pW at 150\,GHz. Therefore, the power from temperature fluctuations is at the sub percent level even with a relatively low sky loading. \par
We also note that the detectors are not affected by LAT turnarounds. For our scan strategy, we expect a turnaround to occur every $120$\,s, or at a rate of roughly $0.008$\,Hz. In the inset plot of Figure \ref{fig:ASD}, we see no noticeable peak at that frequency, indicating the detectors are agnostic to turnarounds.\par

\begin{figure}[h]
    \centering
    \subfloat[90 GHz map]{
        \includegraphics[trim={0 0 2.95cm 0},clip,width=.5\textwidth]{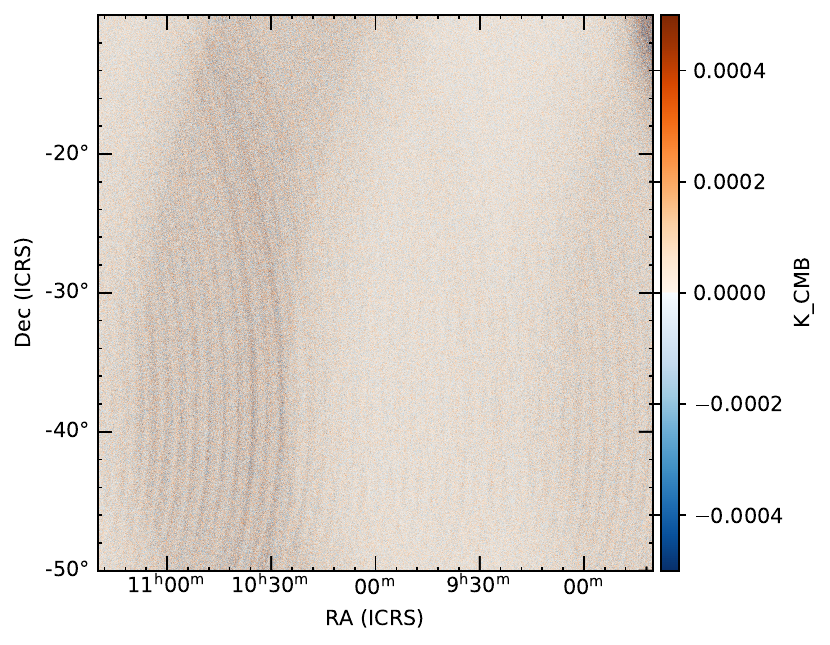}
        }
    \subfloat[150 GHz map]{
        \includegraphics[trim={0 0 2.95cm 0},clip,width=.5\textwidth]{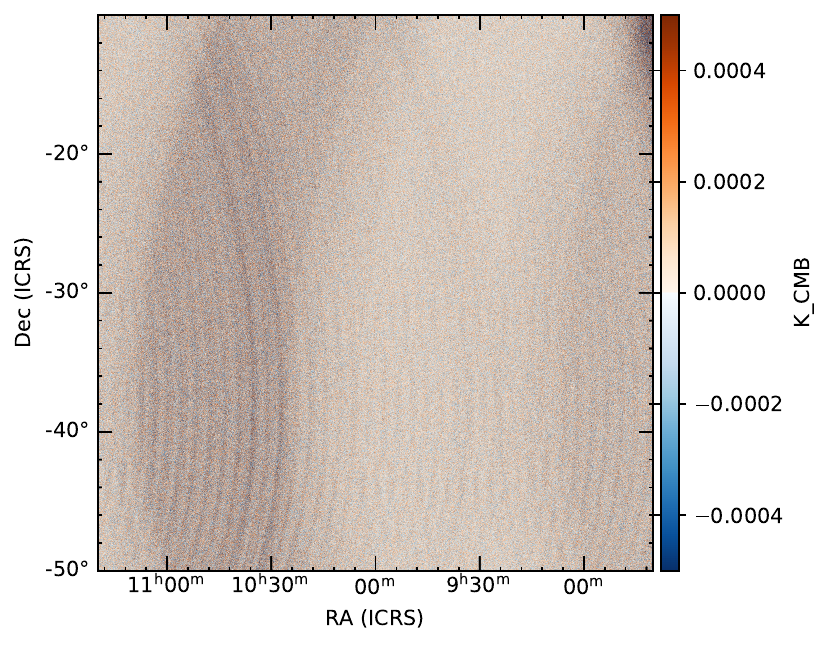}
    }
    \caption{A section of a filter-and-bin map of the dark LAT data, the only filtering done here is subtracting the SVD derived common mode. The mapping speeds reported in Section \ref{sec:LAT} use a map that covers $\sim4$ times the area of those pictured here. The variations in noise levels across these maps are due to uneven coverage, with the arc-like structures corresponding to areas with poor overlap between scans. There is no obvious structure in the maps that point to significant unwanted data making it past our flagging. Figure from \cite{Haridas_2024}.}
    \label{fig:maps}
\end{figure}

Figure \ref{fig:maps} is a dark map we derived from the TODs recorded during LAT scans. Because we are not mapping an actual sky signal, we choose to follow a filter-and-bin mapmaking approach to produce this map, where we simply bin the common mode subtracted TODs into a map.
When collecting on-sky data, we will utilize a maximum likelihood (ML) mapmaker. By using the filter-and-bin method, we can derive the map from the TODs faster on the limited compute capabilities at the high elevation site.
These maps display variations in the noise levels that correspond to areas of uneven sky coverage and have white noise levels in good agreement with our TODs; see \cite{Haridas_2024} for a more complete treatment. \par
\begin{table}[]
\centering
\begin{tabular}{|l|l|l|}
\hline
\textbf{Configuration} & \textbf{90 GHz} & \textbf{150 GHz} \\ \hline\hline
Capped at 4K           & 20.13           & 14.19             \\ \hline
Open to 10K Sky        & 23.44           & 16.92             \\ \hline
\end{tabular}
\caption{K/pW conversions from BoloCalc using SO baseline values~\citep{Hill_2018}. For example, for the 90\,GHz detectors, we expect the sky signal of 10\,K to result in 0.43\,pW of power at the detectors. This is well-below the saturation power ($P_{sat}$) of SO TES, which are designed to be 2.62\,pW for 90\,GHz detectors and 7.18\,pW for 150\,GHz detectors.}
\label{table:bolocalc}
\end{table}

Following dark testing of two MF OTs, we installed four more OTs into the LATR (2 MF and 2 UHF). In anticipation of mirror installation, we removed any metal blanks installed at the cryogenic stages. Finally, during site integration, we noticed imperfections in several of the DSIRs that were meant to be installed in the LATR. This led us to replace all DSIRs with several layers of radio-transparent multi-layer insulation (specifically Zotefoam). At 300\,K (directly behind the window), we installed two $1/8$" layers, one 1" layer, seven more $1/8$" layers. At 80\,K, we installed three more $1/8$" layers. Finally, at 40\,K, we installed two $1/8$" layers. Zotefoam was selected to replace the filters for spectral and thermal reasons~\citep{choi_2013,Ade_2022}. After this switch, we found the thermal loading values below 40\,K to be consistent with prior configurations, but we noticed a larger effect at the 40\,K and 80\,K stages. Specifically: we found $\sim$47\,W at the 80\,K stage, $\sim$65\,W at the 40\,K stage, and $\sim$0.7\,W at the 4\,K stage. The 1\,K and 100\,mK stages were agnostic to this change (we estimate 6.5\,$\mu$W of loading per OT at 100\,mK). Comparing these values to those from prior cooldowns (see Tables \ref{table:80k_loading} and \ref{table:40k_loading}), we note that there is far more loading at 40\,K and less loading at 80\,K in this current configuration. We are currently investigating this change further. Because we do not see a large change in thermal power at the 100\,mK stage, we are comfortable with moving forward using this configuration. Initial Fourier transform spectroscopy indicates that the switch to Zotefoam filters has no impact to MF bandpass, with further analysis on the performance of the UHF OTs underway. Additionally, we report on the performance of the recently installed UFMs in \cite{satterthwaite_2024}. \par

\pagebreak
\section{Conclusion and Future Plans} \label{sec:conclusion}
The Large Aperture Telescope Receiver (LATR) has been integrated into the Simons Observatory (SO) Large Aperture Telescope (LAT) in the Atacama Desert of Chile. The LATR's cryogenic stages have been shown to meet the desired thermal performance requirements in both the laboratory and the LAT. We also show that the thermal performance of the cold optical elements meets its requirement as well. Additionally, the LATR's readout and detector chains have been fully characterized in a dark configuration, and we expect that the detector performance will be suitable for on-sky observations. \par
We expect the LAT's mirrors will be completed and installed in 2025. At the time of writing, there are six OTs (four MF, two UHF) installed in the LATR in Chile~\citep{Haridas_2024,satterthwaite_2024}. The final seven OTs are undergoing development, and are expected to be installed by 2027. In preparation for this final installation, we are exploring minor upgrades to the LATR, in order to improve cryogenic performance. First, as mentioned in Section \ref{sec:LAT}, we replaced our DSIR filters with Zotefoam filters. We will continue to investigate the optical and cryogenic effects of this change. Additionally, previous studies have found that the thermal straps employed in the LATR are not qualified for sub-kelvin performance~\citep{Dhuley_2020}. While the vendor who produced the LATR's straps press-welds the copper braids, we believe that electron-beam welding or press welding this interface may greatly improve their performance. We are currently investigating the feasibility and efficacy of these upgrades. \par
%%%%%%%%%%%%%%%%%%%%%%%%%%%%%%%%%%%%%%%%%%%%%%%%%%%%%%%%%%%%%%%%%%
\section{Acknowledgements}
This work was supported in part by a grant from the Simons Foundation (Award \#457687, B.K.). This work was supported by the National Science Foundation (UEI GM1XX56LEP58).

\section{CR\textsc{edi}T Roles}
We list here the roles and contributions of the authors according to the Contributor Roles Taxonomy (CRediT)\footnote{\url{https://credit.niso.org/}}. \\ 

\noindent\textbf{Tanay Bhandarkar}: Conceptualization (equal), Data curation (equal), Formal analysis (lead), Investigation (lead), Methodology (equal), Resources (equal), Software (equal), Validation (lead), Visualization (equal), Writing - original draft (lead), Writing - review \& editing (lead).\\ 
\textbf{Gabriele Coppi}: Conceptualization (supporting), Investigation (supporting), Methodology (supporting), Resources (supporting), Writing - review \& editing (supporting).\\ 
\textbf{Thomas Satterthwaite}: Conceptualization (supporting), Formal analysis (equal), Investigation (equal), Methodology (supporting), Validation (supporting), Writing - review \& editing (supporting).\\ 
\textbf{Shawn Henderson}: Conceptualization (supporting), Investigation (supporting), Methodology (supporting), Project administration (supporting), Resources (supporting), Software (supporting), Validation (supporting), Writing - review \& editing (supporting).\\ 
\textbf{Federico Nati}: Conceptualization (supporting), Funding acquisition (lead), Investigation (supporting), Methodology (supporting), Resources (supporting), Validation (supporting), Writing - review \& editing (supporting).\\ 
\textbf{Matthew Koc}: Conceptualization (supporting), Investigation (supporting), Methodology (supporting), Resources (supporting), Validation (supporting).\\ 
\textbf{Carlos Sierra}: Conceptualization (supporting), Investigation (supporting), Methodology (supporting), Resources (supporting), Validation (supporting).\\ 
\textbf{Michael Link}: Conceptualization (supporting), Investigation (supporting), Methodology (supporting), Resources (supporting), Validation (supporting), Writing - review \& editing (supporting).\\ 
\textbf{Tammy Lucas}: Conceptualization (supporting), Investigation (supporting), Methodology (supporting), Resources (supporting), Validation (supporting), Writing - review \& editing (supporting).\\ 
\textbf{Michael Vissers}: Conceptualization (supporting), Investigation (supporting), Methodology (supporting), Resources (supporting), Validation (supporting).\\ 
\textbf{Kyuyoung Bae}: Conceptualization (supporting), Investigation (supporting), Methodology (supporting), Resources (supporting), Validation (supporting).\\ 
\textbf{Shannon Duff}: Conceptualization (supporting), Investigation (supporting), Methodology (supporting), Resources (supporting), Validation (supporting), Writing - review \& editing (supporting).\\ 
\textbf{Jenna Moore}: Conceptualization (supporting), Investigation (supporting), Methodology (supporting), Resources (supporting), Validation (supporting), Writing - review \& editing (supporting).\\ 
\textbf{Tran Tsan}: Conceptualization (supporting), Methodology (supporting), Software (supporting), Validation (supporting), Writing - review \& editing (supporting).\\ 
\textbf{Liam Walters}: Conceptualization (supporting), Validation (supporting), Writing - review \& editing (supporting).\\ 
\textbf{Jason Austermann}: Conceptualization (supporting), Resources (supporting), Validation (supporting), Writing - review \& editing (supporting).\\ 
\textbf{Eve Vavagiakis}: Conceptualization (supporting), Resources (supporting), Validation (supporting), Writing - review \& editing (supporting).\\ 
\textbf{Zeeshan Ahmed}: Conceptualization (supporting), Formal analysis (supporting), Funding acquisition (supporting), Investigation (supporting), Methodology (supporting), Project administration (supporting), Resources (supporting), Supervision (supporting), Writing - review \& editing (supporting).\\ 
\textbf{Brian Koopman}: Conceptualization (supporting), Software (supporting), Writing - review \& editing (supporting).\\ 
\textbf{Nicholas Galitzki}: Conceptualization (supporting), Methodology (supporting), Writing - review \& editing (supporting).\\ 
\textbf{Daniel Dutcher}: Conceptualization (supporting), Investigation (supporting), Methodology (supporting), Resources (supporting), Software (supporting), Validation (supporting), Writing - review \& editing (supporting).\\ 
\textbf{Mark Devlin}: Conceptualization (supporting), Funding acquisition (lead), Investigation (supporting), Methodology (equal), Project administration (lead), Resources (lead), Supervision (lead), Writing - review \& editing (supporting).\\ 
\textbf{Robert Thornton}: Conceptualization (supporting), Investigation (supporting), Methodology (supporting), Project administration (supporting), Resources (supporting), Validation (supporting), Writing - review \& editing (supporting).\\ 
\textbf{Simon Dicker}: Conceptualization (supporting), Investigation (supporting), Methodology (supporting), Resources (supporting), Validation (supporting), Writing - review \& editing (supporting).\\ 
\textbf{John Orlowski-Scherer}: Conceptualization (supporting), Formal analysis (supporting), Investigation (supporting), Methodology (supporting), Validation (supporting), Writing - review \& editing (supporting).\\ 
\textbf{Yuhan Wang}: Conceptualization (supporting), Data curation (equal), Formal analysis (supporting), Investigation (supporting), Methodology (supporting), Resources (supporting), Software (supporting), Validation (supporting), Writing - review \& editing (supporting).\\ 
\textbf{Anna Kofman}: Conceptualization (supporting), Formal analysis (supporting), Investigation (supporting), Methodology (supporting), Resources (equal), Software (supporting), Validation (supporting), Visualization (supporting), Writing - review \& editing (supporting).\\ 
\textbf{Saianeesh Haridas}: Conceptualization (supporting), Data curation (equal), Formal analysis (equal), Investigation (supporting), Methodology (equal), Resources (equal), Software (equal), Validation (supporting), Visualization (supporting), Writing - original draft (supporting), Writing - review \& editing (supporting).\\ 
\textbf{Jon Gudmundsson}: Conceptualization (supporting), Resources (supporting), Validation (supporting), Visualization (supporting), Writing - review \& editing (supporting).\\ 
\textbf{Karen Perez Sarmiento}: Conceptualization (supporting), Data curation (supporting), Investigation (supporting), Methodology (supporting), Software (supporting), Validation (supporting), Writing - original draft (supporting).\\ 
\textbf{Mario Zannoni}: Conceptualization (supporting), Funding acquisition (supporting), Investigation (supporting), Methodology (supporting), Resources (supporting), Validation (supporting), Writing - review \& editing (supporting).\\
\textbf{Michele Limon}: Conceptualization (supporting), Investigation (supporting), Methodology (supporting), Project administration (supporting), Supervision (supporting), Validation (supporting), Visualization (supporting).\\ 
\textbf{Suzanne Staggs}: Conceptualization (supporting), Funding acquisition (supporting), Methodology (supporting), Project administration (supporting), Resources (supporting), Supervision (supporting), Validation (supporting).\\ 
\textbf{Michael Niemack}: Conceptualization (supporting), Funding acquisition (supporting), Methodology (supporting), Project administration (supporting), Resources (supporting), Supervision (supporting), Validation (supporting), Writing - review \& editing (supporting).\\
\\

%% For this sample we use BibTeX plus aasjournals.bst to generate the
%% the bibliography. The sample631.bib file was populated from ADS. To
%% get the citations to show in the compiled file do the following:
%%
%% pdflatex sample631.tex
%% bibtext sample631
%% pdflatex sample631.tex
%% pdflatex sample631.tex

\bibliography{main}{}
\bibliographystyle{aasjournal}

%% This command is needed to show the entire author+affiliation list when
%% the collaboration and author truncation commands are used.  It has to
%% go at the end of the manuscript.
%\allauthors

%% Include this line if you are using the \added, \replaced, \deleted
%% commands to see a summary list of all changes at the end of the article.
%\listofchanges

\end{document}